\documentclass[12pt, a4paper]{article}
\usepackage{graphicx}
\usepackage{amssymb}
\usepackage{amsmath}
\usepackage{bm}
\usepackage{comment}
\usepackage{color}
\usepackage[dvipsnames]{xcolor}
\usepackage{cite}
\usepackage{slashed}
\usepackage{subfigure}
\usepackage{epstopdf}            
\usepackage{epsfig}
\renewcommand{\thefootnote}{\fnsymbol{footnote}}
\newcommand{\beq}{\begin{equation}}
\newcommand{\eeq}{\end{equation}}

\newcommand{\la}{\lambda}

\newcommand{\VEV}[1]{{\langle #1 \rangle}}

\definecolor{azur}{rgb}{0.118,0.498,0.796}
\definecolor{darkred}{cmyk}{0,1,1,0.4}
\definecolor{green1}{rgb}{0.21,0.6,0.32}

\usepackage[colorlinks=true, linkcolor=magenta, citecolor=ForestGreen, urlcolor=black]{hyperref}

\begin{document}

\begin{titlepage}

\begin{center}

{\Large
\bf
Light mass window of lepton portal dark matter
}

\vskip 2cm

Shohei Okawa$^{1}$\footnote{\href{mailto:okawa@uvic.ca}{okawa@uvic.ca}}
and
Yuji Omura$^{2}$\footnote{\href{mailto:yomura@phys.kindai.ac.jp}{yomura@phys.kindai.ac.jp}}

\vskip 0.5cm

{\it $^1$ Department of Physics and Astronomy, University of Victoria, \\
Victoria, BC V8P 5C2, Canada}

{\it $^2$
Department of Physics, Kindai University, Higashi-Osaka, Osaka 577-8502, Japan}\\[3pt]

\vskip 1.5cm

\begin{abstract}
We explore a novel possibility that dark matter has a light mass below 1\,GeV 
in a lepton portal dark matter model. 
There are Yukawa couplings involving dark matter, left-handed leptons and an extra scalar doublet in the model.
In the light mass region, dark matter is thermally produced via its annihilation into neutrinos. 
In order to obtain the correct relic abundance and avoid collider bounds, 
a neutral scalar is required to be light while charged scalars need to be heavier than the electroweak scale. 
Such a mass spectrum is realized by adjusting quartic couplings in the scalar potential 
or introducing an extra singlet scalar. 
It turns out that the mass region of 10\,MeV--10\,GeV is almost free 
from experimental and observational constraints. 
We also point out that searches for extra neutrino flux from galactic dark matter annihilations 
with neutrino telescopes are the best way to test our model. 
\end{abstract}

\end{center}
\end{titlepage}

\def\rem#1{ {\bf\textcolor{red}{($\clubsuit$ #1 $\clubsuit$)}}}
\renewcommand{\thefootnote}{\#\arabic{footnote}}
\setcounter{footnote}{0}

\section{Introduction}

A weakly interacting massive particle (WIMP) is one of the most fascinating dark matter (DM) candidates from both theoretical and experimental perspectives. 
The WIMP DM has a mass, typically ranging from ${\cal O}(10)$\,GeV to ${\cal O}(1)$\,TeV, 
and interaction with the Standard Model (SM) particles, 
whose strength is compatible with that of the weak interactions. 
Such a DM candidate is produced in the SM thermal plasma through the DM-SM interaction and 
the produced relic abundance is in agreement with the observed one. 
Besides, the DM scattering with nuclei and annihilation into the SM particles, correlated to the thermal production, would leave observational signals, motivating a lot of experimental efforts. 
The DM experiments, however, have not observed any signals, 
so that the WIMP DM scenario often suffers from the stringent constraints. 

The direct detection experiments, for instance, have set the leading bounds on the DM scattering with nuclei. 
The XENON1T collaboration announces that the upper bound on the
cross section of the elastic scattering
reaches ${\cal O} (10^{-46})$ cm$^2$ if the DM mass is ${\cal O}(100)$\,GeV~\cite{Aprile:2018dbl}. 
This sensitivity is sufficient to test even simplified DM models
where DM does not couple with light quarks at tree level. 
Thus, we have to think out the way to avoid the constraint in the WIMP DM scenario. 
A simple idea is to consider a sub-GeV DM mass. 
In such a region, the direct detection loses the sensitivity due to the detection threshold, 
and the constraint is drastically relaxed. 

Such a sub-GeV DM scenario is, on the other hand, severely constrained from indirect searches, if DM is thermally produced.
For example, it is well-known that the DM annihilation, 
that produces electrons or photons in the end, 
increases the ionization fraction in the post-recombination era and, in turn, disturbs 
the anisotropies of the Cosmic Microwave Background (CMB). 
The resulting upper limit on the annihilation cross section (per DM mass) indicates 
that if the $s$-wave annihilation into charged particles is responsible 
for the thermal production, DM mass is restricted to be heavier than 10\,GeV~\cite{Slatyer:2015jla,Leane:2018kjk}. 
Besides, the Fermi collaboration searches for signals of the DM annihilation in the galactic space. 
The analysis of the latest 11-year data from dwarf galaxies disfavors 
the DM candidate with the mass of several GeV--${\cal O}$(100)\,GeV 
that has the thermal relic cross section for various annihilation modes~\cite{Hoof:2018hyn}. 

Given the above experimental status of the WIMP scenario, 
it may be natural to discuss models with less constrained DM interactions, e.g. interactions with neutrinos. 
In Refs.~\cite{Primulando:2017kxf,Campo:2017nwh}, indeed, the DM production via the interaction with the neutrinos 
has been studied in non-renormalizable effective models. 
While the authors extensively study experimental constraints and DM signals in the light mass region, 
there is a difficulty in embedding such models into renormalizable models. 
When DM couples to neutrinos, DM generally couples to charged leptons as well, to respect the gauge symmetry.
Then, DM (or a mediator of the interaction) is easily coupled to the ordinary matter, 
so that the model eventually suffers from the constraints from the DM experiments. 
Thus, the class of viable models that incorporate the sizable DM-neutrino interactions is limited. 
Feasible renormalizable models, for example, include gauged $U(1)_{L_\mu-L_\tau}$ extensions~\cite{Foldenauer:2018zrz,Bernreuther:2020koj,Asai:2020qlp} and neutrino portal models~\cite{Batell:2017cmf,McKeen:2018pbb,Blennow:2019fhy}. 

In this paper, 
we propose another simple renormalizable model of light DM that dominantly interacts with neutrinos. 
In this model, DM has Yukawa couplings to left-handed leptons introducing an extra scalar doublet. 
This kind of model is called the lepton portal DM model~\cite{Bai:2014osa,Chang:2014tea,Kawamura:2020qxo}. 
In Ref. \cite{Kawamura:2020qxo}, the authors classify the lepton portal DM models,
 according to the spin of DM and the electroweak charge of the leptonic mediator. 
One of them has the same structure as the model discussed in this paper.
In Ref. \cite{Kawamura:2020qxo}, no mass splitting between the neutral and charged components of the extra scalar is assumed. Then, we find that the sub-GeV DM cannot be thermally produced 
due to the collider bounds and the perturbativity violation. 
In this paper, therefore, we explore the sub-GeV mass window assuming the mass splitting is large. 
In order to obtain the correct relic abundance and avoid collider bounds, 
the neutral scalar is required to be lighter than ${\cal O}$(10)\,GeV, 
while the charged scalar needs to be heavier than the electroweak (EW) scale. 
It is interesting that the lepton portal model already has the capacity of realizing such a mass spectrum in the scalar potential, without additional extension. 
We have several scalar quartic couplings as free parameters and 
can make the ${\cal O}$(100)\,GeV mass splitting by adjusting their values. 
As a result of the specific mass spectrum, the model sharply predicts the ${\cal O}(1)$ quartic couplings and characteristic signals of the 125\,GeV Higgs boson. 
If future collider experiments improve the measurement of the 125\,GeV Higgs and the search for the extra charged scalar, 
this model can be tested explicitly. 
In most part of this paper, we focus on the most economical model with one extra doublet scalar, 
while we also extend the model introducing an additional singlet scalar. 
Such an extension helps to relax the strong limitations on 
the scalar quartic couplings and the Higgs signal strength, and then resolves the collider constraints. 
We note that a sufficiently light neutral scalar can be realized in both models.

In DM physics, we figure out that DM is correctly produced via its annihilation into active neutrinos 
in the DM mass region of 10\,MeV--10\,GeV. 
Even future direct detection experiments, CMB observation and indirect gamma-rays searches 
will not be able to cover this mass region. 
We suggest that neutrino flux observation with future neutrino telescopes is the best way to 
test this DM candidate. 
Although the current best limit from the Super-Kamiokande (SK) experiment 
is one order of magnitude larger than the thermal relic cross section, 
the future neutrino observatories, such as Hyper-Kamiokande (HK), DUNE and JUNO, will have enough sensitivity to probe it. 

This paper is organized as follows. 
In Sec.~\ref{sec2}, we detail the model setup and show that the annihilation cross section can be sufficiently large if and only if one neutral scalar mediator is lighter than ${\cal O}$(10)\,GeV. 
In Sec.~\ref{sec3}, the experimental constraints on the extra scalars are summarized and, 
in Sec.~\ref{sec4}, the DM constraints and signals are investigated. 
Section \ref{sec5} is devoted to summary.

%%%%%%%%%%%%%%%%%%%%%%%%%%%%%%%%%%%%%%%%%%%%%%%%%%%%%%%%%%%
\begin{table}[t]
\begin{center}
\begin{tabular}{ccccccc}
\hline
\hline
Fields & spin   & ~~$SU(3)$~~ & ~~$SU(2)_L$~~  & ~~$U(1)_Y$~~   &   ~~$U(1)_L$~~   &   ~~$Z_2$~~    \\ \hline  
  $Q_L^{i }$  & $1/2$ & ${\bf 3}$         &${\bf2}$    &   $\frac{1}{6}$  &   0  &   +  \\  
   $u_R^{i }$  & $1/2$&  ${\bf 3}$        &${\bf1}$&  $\frac{2}{3}$  &   0  &   +      \\ 
     $d_R^{i }$  & $1/2$&  ${\bf 3}$        &${\bf1}$&  $-\frac{1}{3}$  &   0  &   +     \\   \hline
     $\ell^i_L$ & $1/2$ &  ${\bf 1}$        &${\bf 2}$ &   $-\frac{1}{2}$  &   1  &   +    \\ 
      $e^i_R$ & $1/2$ &  ${\bf 1}$        &${\bf 1}$ &   $-1$  &   1  &   +       \\   \hline
     $\psi_L$ & $1/2$ &  ${\bf 1}$        &${\bf 1}$ &   $0$  &   1  &   $-$     \\  
     $\psi_R$ & $1/2$ &  ${\bf 1}$        &${\bf 1}$ &   $0$  &   1  &   $-$     \\  \hline
     $\Phi$ & $1$ & ${\bf 1}$         &${\bf2}$ &   $\frac{1}{2}$  &   0  &   +      \\  
          $\Phi_\nu$ & $1$ & ${\bf 1}$         &${\bf2}$ &   $\frac{1}{2}$  &   0  &   $-$        \\ 
                       \hline \hline
\end{tabular}
\end{center}
\caption{The matter content of the model. $i$ denotes the flavors: $i=1, \, 2, \, 3$. 
$U(1)_L$ and $Z_2$ are global symmetry.}
 \label{table1}
\end{table} 
%%%%%%%%%%%%%%%%%%%%%%%%%%%%%%%%%%%%%%%%%%%

\section{Lepton Portal Model with fermion DM}
\label{sec2}

We introduce an extended SM with an extra scalar doublet and a fermion DM candidate.
The matter content is shown in Table \ref{table1}. 
$Q_L^{i }$, $u_R^{i }$ and $d_R^{i }$ denote left-handed and right-handed quarks, respectively.
$\ell^i_L$ and $e^i_R$ are left-handed and right-handed leptons that carry lepton numbers defined by
the global symmetry, $U(1)_L$. A Dirac fermion, $\psi=(\psi_L, \psi_R)^T$, is introduced as a DM candidate 
that is SM gauge singlet, but charged under $U(1)_L$. 
We have two EW doublet scalar fields, $\Phi$ and $\Phi_\nu$. 
The former denotes the SM Higgs field that breaks the EW symmetry. 
The other scalar $\Phi_\nu$ has the same charge as that of $\Phi$, 
but we assume it does not develop the vacuum expectation value (VEV). 
A $Z_2$ symmetry is imposed on the model: the SM fields are $Z_2$-even and 
the extra fields, $\Phi_\nu$ and $\psi$, are $Z_2$-odd. 
In the rest of this section, we review the couplings relevant to the phenomenology.

\subsection{Lepton portal coupling}
In our model, the DM candidate, $\psi$, couples to the left-handed leptons via the Yukawa couplings
involving $\Phi_\nu$, 
\beq
\label{Yukawa}
- {\cal L}_{\ell} = y_\nu^i \, \overline{\ell_L^i} \, \widetilde{\Phi_\nu} \,  \psi_R +h.c.,
\eeq
with $\widetilde{\Phi_\nu} = i\sigma_2 \Phi_\nu^*$. 
The couplings $y_\nu^i$ dominantly contribute to the annihilation of $\psi$.
This kind of model is called the lepton portal DM model, 
and the phenomenological aspects are summarized in Ref. \cite{Ibarra:2015fqa,Kawamura:2020qxo}, 
assuming the DM mass is more than 100\,GeV. 
In this paper, we explore the possibility that the DM mass, $m_\psi$, is less than 1\,GeV.

\begin{figure}[t]
\centering
\includegraphics[viewport=150 600 400 760, clip=true, scale=0.65]{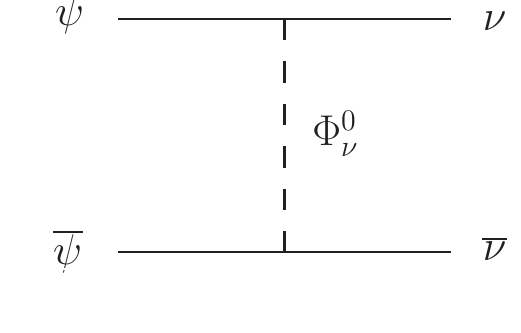} \hspace{1cm}
\includegraphics[viewport=150 600 400 760, clip=true, scale=0.65]{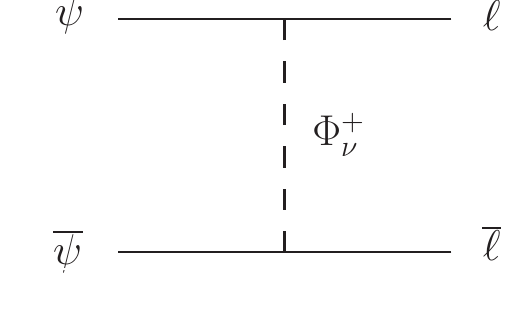}
\caption{DM annihilation processes into neutrinos and charged leptons. 
We denote the neutral and charged components of $\Phi_\nu$ as $\Phi_\nu^0$ and $\Phi_\nu^+$, respectively.}
\label{fig;ann}
\end{figure}

It is illuminating to know what condition is needed to realize the light lepton portal DM. 
To figure it out, 
we consider a simple case that the charged and neutral components of $\Phi_\nu$ have a common mass. 
In this case, the leading DM annihilation proceeds via the $t$-channel $\Phi_\nu$ exchanging (Fig.~\ref{fig;ann}) and the cross section is estimated as 
\begin{equation}
(\sigma v) %\simeq \frac{y_\nu^4 m_\psi^2}{4\pi m_{\Phi_\nu}^4} 
\simeq 10^{-27} \, {\rm cm^3/s} 
\times y_\nu^4 \left(\frac{m_\psi}{\rm GeV}\right)^2 \left(\frac{100\,{\rm GeV}}{m_{\Phi_\nu}}\right)^4 ,
\end{equation}
where we assume $m_\psi \ll m_{\Phi_\nu}$, omit the flavor index, and only keep the partial $s$-wave. 
Since the charged scalar should be heavier than 100\,GeV due to collider bounds, 
the cross section is too small to achieve the canonical value $\sim 10^{-25}\,{\rm cm^3/s}$ for the sub-GeV Dirac DM~\cite{Steigman:2012nb,Saikawa:2020swg}. 
This forces us to introduce a sizable mass splitting between the charged and neutral components of $\Phi_\nu$. 
As discussed in the next subsection, the lepton portal DM model can realize the mass splitting, 
not introducing any additional degrees of freedom. 

%===================
%---------

\subsection{Scalar potential}

The potential for the scalar fields is given by 
\begin{align}
V & = m_1^2 (\Phi^\dagger \Phi) 
   + m_2^2 (\Phi_\nu^\dagger \Phi_\nu) 
   + \lambda_1 (\Phi^\dagger \Phi)^2 
   + \lambda_2 (\Phi_\nu^\dagger \Phi_\nu)^2 \nonumber\\
   & \quad + \lambda_3 (\Phi^\dagger \Phi) (\Phi_\nu^\dagger \Phi_\nu)
   + \lambda_4 (\Phi^\dagger \Phi_\nu) (\Phi_\nu^\dagger \Phi) 
   + \frac{1}{2} \lambda_5 [ (\Phi^\dagger \Phi_\nu)^2 + h.c. ]  .
%   &\quad  + \frac{1}{2} \lambda_{\Phi S} S^2 (\Phi^\dagger \Phi) + \frac{1}{2} \lambda_{\Phi_\nu S} S^2 (\Phi_\nu^\dagger \Phi_\nu)   + \frac{\lambda_S}{4 !} S^4+A \, S \Phi^\dagger \Phi_\nu+ h.c. .
\end{align}
The Higgs sector is the same as that of the inert two Higgs doublet model~\cite{Barbieri:2006dq,Belyaev:2016lok,Belyaev:2018ext}. 
The potential is bounded from below if and only if 
\begin{equation}
\lambda_1 > 0, \quad \lambda_2 > 0, \quad \lambda_3 + 2\sqrt{\lambda_1\lambda_2} > 0 ,\quad 
\lambda_3 + \lambda_4 + \lambda_5 + 2\sqrt{\lambda_1\lambda_2} > 0 .
\label{eq;BB}
\end{equation}
Since there are multiple minima, we keep the inert vacuum from being the unstable false vacuum. 
This is achieved if the inequality, 
\begin{equation}
\lambda_4 + \lambda_5 < 0 .
\label{eq;NCB}
\end{equation}
is satisfied~\cite{Belyaev:2016lok}. 

Let us denote the physical modes as 
\begin{equation}
\Phi = \begin{pmatrix} G^+ \\ \frac{1}{\sqrt{2}} (v+h+iG^0) \end{pmatrix} ,\quad 
\Phi_\nu = \begin{pmatrix} H^+ \\ \frac{1}{\sqrt{2}} (H+iA) \end{pmatrix} .
\end{equation}
where $G^0$ and $G^\pm$ are the NG modes eaten by the $Z$ and $W^\pm$ bosons, respectively.
$v$ denotes the non-vanishing VEV: $v \simeq 246$ GeV.
The masses of the physical scalars are expressed in terms of the quartic couplings and the Higgs VEV: 
\begin{align}
m_h^2 & = 2 \lambda_1 v^2 , \label{eq;Mh} \\
m_{H^+}^2 & = m_2^2 + \frac{\lambda_3 v^2}{2} , \label{eq;MHC} \\
m_A^2 & = m_{H^+}^2 + \frac{(\lambda_4-\lambda_5) v^2}{2} , \label{eq;MA} \\
m_H^2 & = m_{H^+}^2 +\frac{(\lambda_4+\lambda_5) v^2}{2} . \label{eq;MH}
\end{align}
In this paper, we consider $H$ as the lightest inert scalar without loss of generality. 
From the above equations, we see that $H$ can be light by setting  
\begin{equation}
\lambda_5 < 0 , \quad |\lambda_5|= {\cal O}(1) ,
\end{equation}
while keeping $H^+$ and $A$ heavy. 
The scalar quartic couplings and thus the scalar mass spectrum are further restricted 
by various collider bounds and precision observables,
whose detail is discussed in the next section. 
As will be seen in Sec.~\ref{sec;precision}, for example, 
we set $m_{H^+}$ around $m_A$, 
to avoid the constraints from the EW precision observables, especially from the $T$ parameter. 
This is realized by choosing 
\begin{equation}
\lambda_4 \approx \lambda_5 .
\end{equation}
Then, the global minimum condition Eq.(\ref{eq;NCB}) is automatically satisfied since $\lambda_5<0$. 
In Sec.~\ref{sec;hinv}, we will also show that $|\lambda_3+\lambda_4+\lambda_5|$ should be very small 
due to the constraints from the Higgs invisible decay, 
so that we expect that the fourth condition of Eq.(\ref{eq;BB}) will always be satisfied in the phenomenological study. 
Note that $|\lambda_3+\lambda_4+\lambda_5| \ll 1$ means 
\begin{equation}
\lambda_3 \approx -(\lambda_4+\lambda_5) = {\cal O}(1) .
\end{equation}
Since the coupling $\lambda_3$ contributes to the Higgs decay into $\gamma\gamma$ via the charged scalar loop, a deviation of the Higgs signal strength will be predicted in this setup.

\section{Constraints on the extra scalars}
\label{sec3}

In this section, we discuss the prime constraints on the extra scalars. 
We will find that several parameters are almost fixed 
when $H$ is lighter than 10\,GeV. 
This results in the strong predictions of some observables.

\subsection{Oblique corrections}
\label{sec;precision}

New particles with the EW charges contribute to the vacuum polarization of the weak bosons and photon. 
The corrections are rendered in the $S$, $T$ and $U$ parameters. 
The $S$ and $T$ parameters in this model are expressed in terms of the extra scalar masses. 
The $S$ parameter is given by~\cite{Barbieri:2006dq,Belyaev:2016lok,Belyaev:2018ext} 
\begin{equation}
S = \frac{1}{72\pi} \frac{1}{(x_2^2-x_1^2)^3} 
\left[ x_2^6 f_a(x_2) - x_1^6 f_a(x_1) + 9 x_2^2 x_1^2 (x_2^2 f_b(x_2) - x_1^2 f_b(x_2)) \right] ,
\end{equation}
where $x_1 = m_H/m_{H^+}$, $x_2=m_A/m_{H^+}$ and 
\begin{equation}
f_a(x) = -5+12\log(x), \quad 
f_b(x) = 3-4\log(x) .
\end{equation}
The $T$ parameter is 
\begin{equation}
T = \frac{1}{32\pi^2 \alpha v^2} 
\left[ f_c(m_{H^+}^2, m_A^2) + f_c(m_{H^+}^2, m_H^2) - f_c(m_A^2, m_H^2) \right] ,
\end{equation}
where 
\begin{equation}
f_c(x,y) = \frac{x+y}{2} - \frac{xy}{x-y} \log\left(\frac{x}{y}\right) .
\end{equation}
It follows that $T=0$ when $m_{H^+} = m_A$. 
The contribution to the $U$ parameter is small, not providing a strong constraint. 
Hence, we neglect it in this paper. 

\begin{figure}[t]
\includegraphics[width=0.5\textwidth]{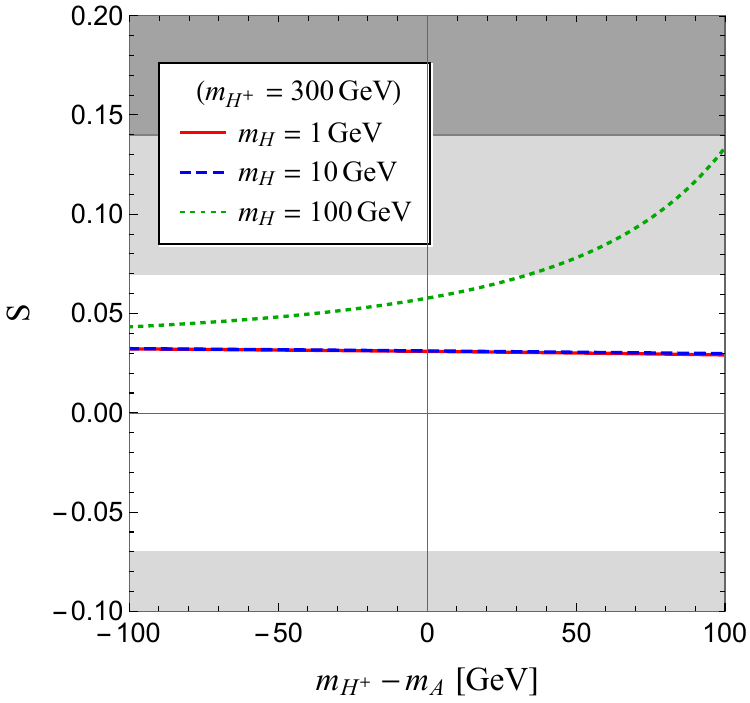}
\includegraphics[width=0.5\textwidth]{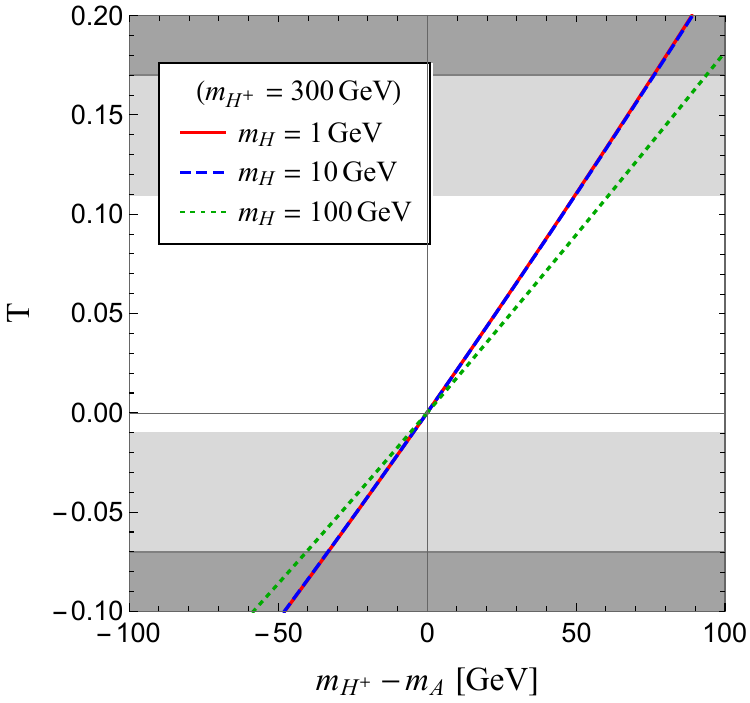}
\caption{The contributions to the $S$ and $T$ parameters in our model. 
In both panels, the red, blue and green lines correspond to $m_H = 1,\,10,\,100$\,GeV, respectively.
The charged scalar mass is fixed at $m_{H^+}=300$\,GeV. 
The light gray and dark gray regions are excluded by the current measurement with $1\sigma$ and $2\sigma$, respectively: $S = 0.00 \pm 0.07$ and $T =  0.05 \pm 0.06$ with $U=0$ \cite{Zyla:2020zbs}.}
\label{fig;STU}
\end{figure}

In Fig.~\ref{fig;STU}, 
we show the contributions to the $S$ and $T$ parameters from the extra scalars 
with the fixed charged scalar mass $m_{H^+}=300$\,GeV. 
The light gray and dark gray regions are excluded by the current measurement 
with $1\sigma$ and $2\sigma$, respectively: 
$S = 0.00 \pm 0.07$ and $T =  0.05 \pm 0.06$ with $U=0$ \cite{Zyla:2020zbs}. 
We see that the limit of the $S$ and $T$ parameters is escaped for a light $H$ 
if the mass splitting of $H^+$ and $A$ is within 50\,GeV. 
Here, we add that the constraint from the $S$ and $T$ parameters requires a small mass splitting between $H^+$ and $A$, 
and hence may require a tuning between $\la_4$ and $\la_5$. 
With $m_{H^+} = 300$\,GeV, for example, we find $\la_4 - \la_5 \simeq - 0.9$ (0.2) with $m_{H^+}-m_A=50$\,GeV (10\,GeV). 
Since $\la_4 \simeq \la_5 \simeq {\cal O}(1)$ in our setup, 
the tuning between $\la_4$ and $\la_5$ is moderate and is not worse than 10\% level. 
As we will see in the next subsection, however, 
the constraint from the Higgs invisible decay requires \% level fine-tuning which is a little more serious.

%===================
%===================

\subsection{Higgs invisible decay}
\label{sec;hinv}

The interaction between $H$ and the SM Higgs boson $h$ is given by 
\begin{equation}
V \supset \frac{\la_{345}}{4} (2 vh + h^2) H^2 ,
\end{equation}
with $\la_{345}=\la_3+\la_4+\la_5$. 
If $H$ is lighter than $m_h/2$, 
the SM Higgs can decay into a pair of $H$ via the coupling $\la_{345}$. 
Then, since $H$ cannot decay to charged particles, 
it contributes to the invisible decay of the Higgs boson. 
The partial decay width is given by 
\begin{equation}
\Gamma(h \to HH) = \frac{|\la_{345}|^2 v^2}{32 \pi m_h} \sqrt{1-\frac{4m_H^2}{m_h^2}} .
\end{equation}
The branching ratio of the Higgs invisible decay is expressed by 
\begin{equation}
{\rm Br}_{\rm inv} = \frac{\Gamma(h \to HH)}{\Gamma_{h, {\rm SM}}+\Gamma(h \to HH)} ,
\end{equation}
where $\Gamma_{h, {\rm SM}}\simeq 4.1$\,MeV is the total decay width of the Higgs boson in the SM. 
Thus, there is an upper limit on the coupling, 
\begin{equation}
|\la_{345}| < \left( \frac{ 32\pi m_h \Gamma_{h, {\rm SM}} }{ v^2 ({\rm Br}_{\rm inv}^{-1} - 1) \sqrt{1-\frac{4m_H^2}{m_h^2}} } \right)^{1/2} .
\end{equation}
The current constraint from the Higgs invisible decay width is 
\begin{equation}
{\rm Br}_{\rm inv} < 
\left\{ 
\begin{array}{ll} 
0.13 & \quad \mbox{(ATLAS~\cite{ATLAS:2020cjb})}\\
0.19 & \quad \mbox{(CMS~\cite{Sirunyan:2018owy})}
\end{array}
\right. .
\end{equation}
For reference, we show the prospect of various future collider experiments summarized in \cite{deBlas:2019rxi}
\begin{equation}
{\rm Br}_{\rm inv} < 
\left\{ 
\begin{array}{ll} 
0.019 & \quad \mbox{(HL-LHC)}\\
0.0026 & \quad \mbox{(ILC(250))}\\
0.00024 & \quad \mbox{(FCC)}
\end{array}
\right. .
\end{equation}

\begin{figure}[t]
\includegraphics[width=0.5\textwidth]{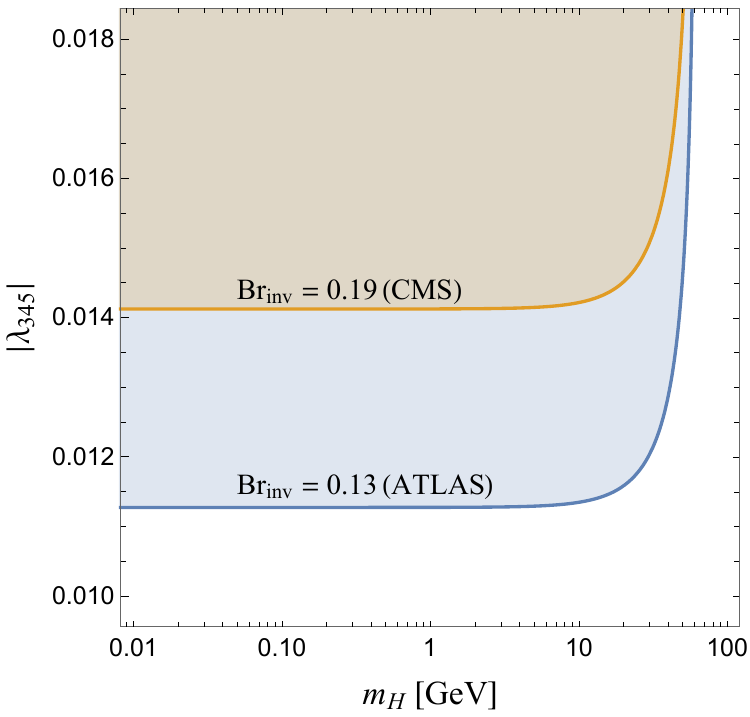}\hspace{0.3cm}
\includegraphics[width=0.5\textwidth]{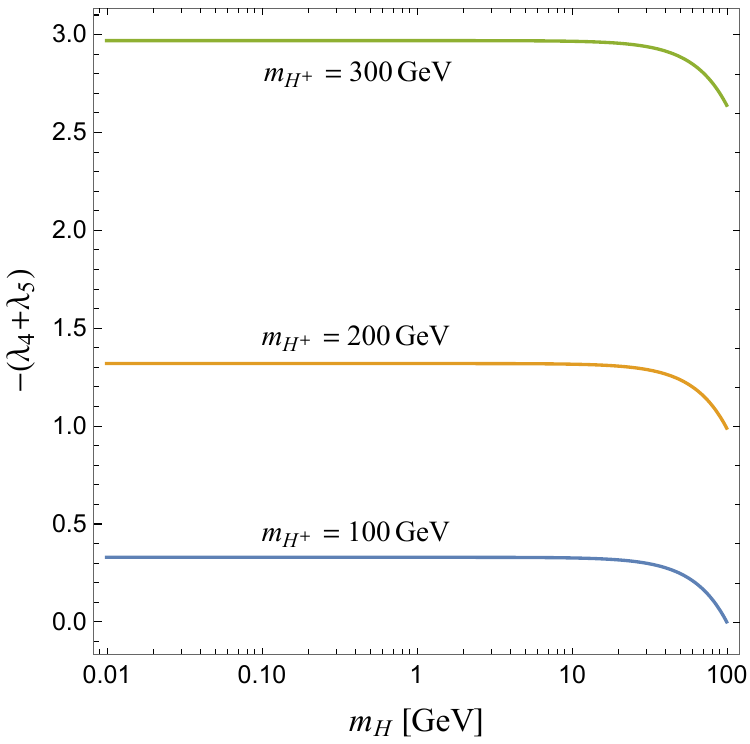}
\caption{(Left) Upper bounds on $|\la_{345}|$ from the Higgs invisible decay. 
The shaded regions are excluded by the current measurement: 
${\rm Br}_{\rm inv} < 0.13$ (ATLAS) \cite{ATLAS:2020cjb} and 0.19 (CMS) \cite{Sirunyan:2018owy}. 
(Right) Three lines show the values of $\la_4+\la_5$ that are determined by Eq.~(\ref{eq;MH}) with $m_{H^+} = 100,\,200,\,300$\,GeV.}
\label{fig;Brinv}
\end{figure}

In Fig.~\ref{fig;Brinv} (left), we show the current constraints on 
the coupling $|\lambda_{345}|$ as a function of $m_H$. 
The shaded regions are excluded by the current measurement of the invisible Higgs decay at ATLAS and CMS. 
We find the upper limit to be $|\la_{345}| \lesssim 0.01$ for $m_H < m_h/2$. 
This means the following relation,
\begin{equation}
\la_3 \approx -(\la_4+\la_5) ,
\end{equation}
should be satisfied at \% level in the light $H$ case. 
Note that $\la_4+\la_5$ is negative for $H$ to be light (see Fig.~\ref{fig;Brinv} (right) and Eq.(\ref{eq;MH})), so that $\la_3$ is positive. 
The sign of $\la_3$ is important in the contribution to the Higgs signal strength as shown in the next subsection. 

%===================
%===================

%===================
\begin{figure}[t]
\centering
\includegraphics[viewport=130 570 430 760, clip=true, scale=0.6]{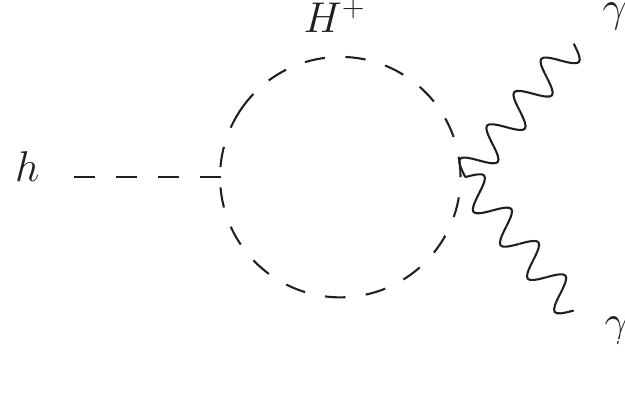} \hspace{1cm}
\includegraphics[viewport=130 570 450 760, clip=true, scale=0.6]{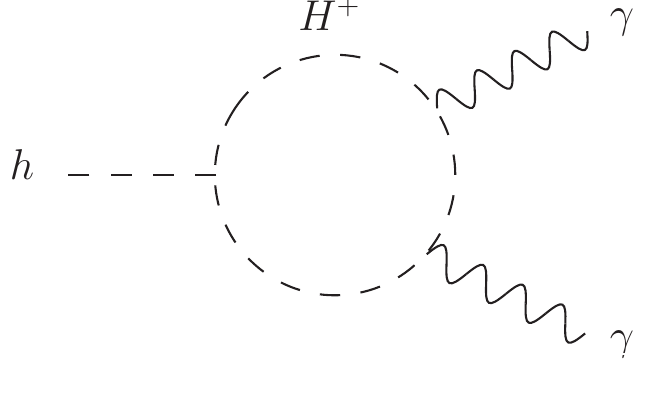}
\caption{The charged scalar contribution to the SM Higgs decay into $\gamma\gamma$.}
\label{fig;hdecay}
\end{figure}
%===================

\subsection{Higgs signal strength $R_{\gamma\gamma}$}

The interaction between the charged scalar and the SM Higgs, 
\begin{equation}
V \supset \frac{\la_3}{2} (2vh+h^2) H^+ H^- ,
\end{equation}
can contribute to the SM Higgs decay into $\gamma\gamma$ at one-loop level (Fig.~\ref{fig;hdecay}). 
The one-loop charged scalar contribution deviates the Higgs signal strength into $\gamma\gamma$, 
which is defined by the following ratio 
\begin{equation}
R_{\gamma\gamma} = \frac{\sigma(pp\to h)}{\sigma(pp\to h)_{\rm SM}} \times \frac{{\rm Br}(h \to \gamma\gamma)}{{\rm Br}(h \to \gamma\gamma)_{\rm SM}} . 
\end{equation}
Since the production cross section is not changed, we find 
\begin{equation}
R_{\gamma\gamma} = \frac{{\rm Br}(h \to \gamma\gamma)}{{\rm Br}(h \to \gamma\gamma)_{\rm SM}} ,
\end{equation}
in this model. 

\begin{figure}[t]
\includegraphics[width=0.5\textwidth]{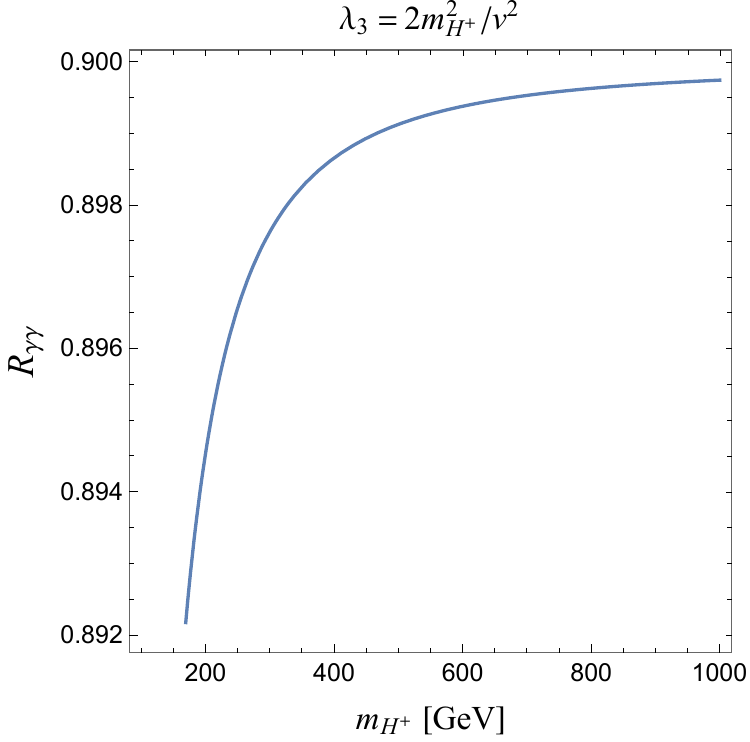}
\includegraphics[width=0.5\textwidth]{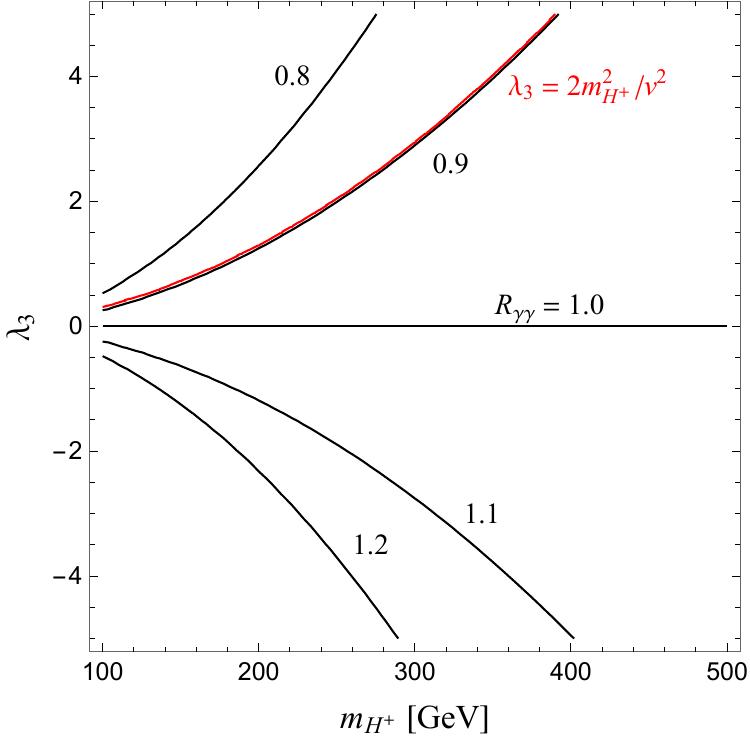}
\caption{(Left) The prediction of the Higgs signal strength $R_{\gamma\gamma}$ in the light $H$ limit. 
(Right) The values of the signal strength $R_{\gamma\gamma}$ in the $(m_{H^+},\, \la_3)$ plane (black lines). 
The prediction in the light $H$ limit is shown with the red line.}
\label{fig;Rgamma}
\end{figure}

Let us evaluate the charged scalar contribution to $R_{\gamma\gamma}$. 
In the inert doublet Higgs model, the partial Higgs decay width into $\gamma\gamma$ is modified as~\cite{Kanemura:2016sos} 
\begin{equation}
\Gamma(h \to \gamma\gamma) = \frac{\alpha^2 m_h^3}{256 \pi^3 v^2} 
\left| \sum_f N_c^{(f)} q_f^2 F_{1/2} (\tau_f) + F_1 (\tau_W) + \frac{\lambda_3 v^2}{2m_{H^+}^2} F_0 (\tau_{H^+}) \right|^2 ,
\label{eq;Rgamma}
\end{equation}
where we defined $F_{\rm spin}(\tau_i)$ and $\tau_i \equiv 4m_i^2/m_h^2$ for a charged particle $i$ 
running in the loop: 
\begin{align}
F_{1} (\tau_i) & = 2 + 3 \tau_i + 3 \tau_i (2-\tau_i) f(\tau_i) ,\\
F_{1/2} (\tau_i) & = -2 \tau_i \left[ 1 + (1-\tau_i) f(\tau_i) \right] ,\\
F_0 (\tau_i) & = \tau_i \left[ 1 - \tau_i f(\tau_i) \right] .
\end{align}
The loop function $f(\tau)$ is 
\begin{equation}
f(\tau) = \left\{ 
\begin{array}{ll} 
\left[ \sin^{-1}(1/\sqrt{\tau} \right)^2 & \quad (\tau \geq 1) \\
-\frac{1}{4} \left[ (\log(\eta_+/\eta_-) - i \pi \right]^2 & \quad (\tau < 1) 
\end{array} \right.
\end{equation}
with 
\begin{equation}
\eta_\pm = 1 \pm \sqrt{1-\tau} .
\end{equation}
Note that the branching ratio only depends on the quartic coupling $\lambda_3$ and the charged scalar mass $m_{H^+}$. 
In Fig.~\ref{fig;Rgamma} (right), we show the signal strength $R_{\gamma\gamma}$ in the $(m_{H^+},\,\lambda_3)$ plane with the black lines. 
It follows that the sign of $\la_3$ determines if the charged scalar contribution is constructive or destructive. 
If $\la_3$ is positive, the charged scalar contribution is destructive to the SM contribution and $R_{\gamma\gamma}$ is smaller than unity, while if $\la_3$ is negative, $R_{\gamma\gamma}$ is larger than unity. 
The signal strengths measured by ATLAS and CMS in LHC Run2 are 
$R_{\gamma\gamma}= 1.02\pm0.14$~\cite{ATLAS:2019jst} and 
$R_{\gamma\gamma}=1.18^{+0.17}_{-0.14}$~\cite{Sirunyan:2018ouh}, respectively. 
These results favor a negative $\la_3$ more than a positive one. 

%===================
%===================

Let us see the prediction of the light $H$ scenario. 
For our purpose, we consider $m_H < 10$\,GeV here. 
As pointed out above, there is the strong relation, $\la_3 \approx -(\la_4+\la_5)$, from the invisible Higgs decay. 
In addition, it is seen in Fig.~\ref{fig;Brinv} (right) that $\la_4 + \la_5$ is almost constant with a fixed $m_{H^+}$ in the light $H$ regime. 
Given these facts, we find that $\la_3$ and $\la_4+\la_5$ are expressed solely by $m_{H^+}$ up to ${\cal O}(m_H^2/v^2)$ corrections:
\begin{equation}
\la_3 \simeq - (\la_4 + \la_5) \simeq \frac{2m_{H^+}^2}{v^2} + {\cal O}\left(\frac{m_H^2}{v^2}\right).
\label{eq;appla3}
\end{equation}
Then, we can remove $\la_3$ from Eq.(\ref{eq;Rgamma}) and 
express $R_{\gamma\gamma}$ as a function of $m_{H^+}$. 
We show the approximate prediction of $R_{\gamma\gamma}$ 
in Fig.~\ref{fig;Rgamma} (left) with $\la_3$ fixed at the value of Eq.(\ref{eq;appla3}). 
In Fig.~\ref{fig;Rgamma} (right), we also plot the same prediction 
in the ($m_{H^+},\,\la_3$) plane with the red line. 
The signal strength is about 10\% smaller than the SM value since $\la_3$ takes a large positive value. 
This is consistent with the ATLAS result while indicates a mild $2.0\,\sigma$ deviation from the CMS one.
Future precise measurement of the Higgs boson will further improve the uncertainties and may discover the deviation or exclude this scenario, but the light $H$ regime is viable at this moment.
Note that the above conclusion is independent of the mass splitting between $H^+$ and $A$, because the mass splitting only depends on $\la_4-\la_5$, while the above discussion only relies on $\la_3$ and $\la_4+\la_5$.

%===================
\begin{figure}[t]
\centering
\includegraphics[viewport=160 560 430 780, clip=true, scale=0.65]{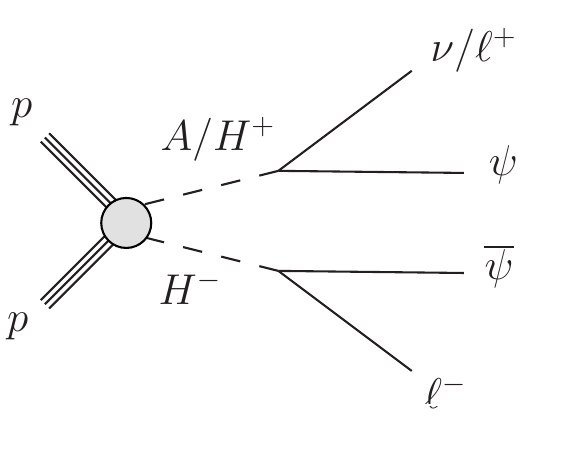} \hspace{1cm}
\includegraphics[viewport=160 560 430 780, clip=true, scale=0.65]{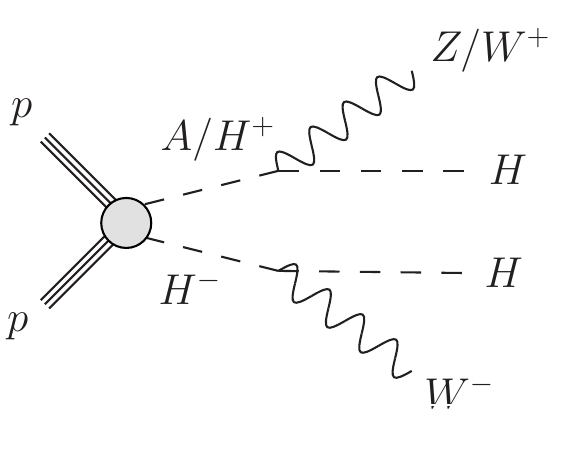}
\caption{Example diagrams for the electroweak production of the extra scalars and DM signals at the LHC.}
\label{fig;collider}
\end{figure}
%===================

\subsection{Collider searches}

The extra scalars interact with quarks via the EW interaction, 
so that they can be produced at the LHC.
The proton-proton collision produces $H^+H^-$, $AH^{\pm}$, $HH^{\pm}$ and $H$$A$.
The production cross section is large enough to test this model, if the scalars are around the EW scale.
The charged scalar, $H^\pm$, decays to $\ell^\pm$$\psi$ ($\ell=e, \, \mu, \, \tau$) and/or $W^\pm$$H$. 
The main decay mode depends on the size of the Yukawa coupling, $y^i_\nu$.
If the Yukawa coupling is large, $H^\pm$ dominantly decays to $\ell^\pm \psi$, 
so that the process, $pp \to H^+ H^-$, leads 
a signal of $\ell^+ \ell^-$ plus large transverse missing energy (Fig.~\ref{fig;collider} (left)). 
Such a characteristic signal is well studied to search for supersymmetric particles and 
the bound on superpartners of leptons, namely sleptons, can be applied to our model.

If $H^\pm$ dominantly decays to muon or electron and $\psi$, 
the lower bound of $m_{H^\pm}$ can be estimated as $m_{H^\pm} \gtrsim 550$\,GeV~\cite{Aad:2019vnb}. 
This bound is too strong to realize the light $H$ without any conflict with the perturbativity of $\lambda_3$. 
If the main decay mode is $H^\pm \to \tau^\pm \psi$, however, the bound is much weak. 
Following Ref.~\cite{Aad:2019byo}, we find the lower bound of $m_{H^\pm}$ to be $m_{H^\pm} \gtrsim 310$\,GeV. 
Note that these bounds are further relaxed, if the Yukawa coupling is small 
and $H^\pm$ dominantly decays to $W^\pm$$H$ (Fig.~\ref{fig;collider} (right)). 
In our model, the branching ratio of
$H^\pm \to \ell^\pm\psi$ is less than $1/2$ 
if $y^i_\nu \leq 1$ and $m_{H^\pm} \geq 300$ GeV.
Thus, the bound from $2 \ell$ plus missing energy is not so strong as discussed above. 

The signal originated from the $H^\pm A$ production possibly brings the strongest constraint to our model. 
The scalar production is followed by $H^\pm \to W^\pm H$ and $A \to ZH$, 
and then the signal is $ZW^\pm $ plus missing energy, since $H$ only decays to $\nu \psi$.
This signal is studied in the neutralino-chargino search \cite{ATLAS:2020ckz}. 
We estimate the production cross section of $H^\pm A$, using {\tt MadGraph5\_aMC@NLO} \cite{Alwall:2014hca}, and compare it with the neutralino and chargino production. 
Based on Ref. \cite{ATLAS:2020ckz}, the lower bound on the chargino mass is about 650 GeV, 
that is translated into the bound on $m_{H^\pm}$, assuming the acceptance is the same as that of the chargino-neutralino search. We find that the charged scalar mass should satisfy $m_{H^\pm}\gtrsim 250$\,GeV, which we use as the collider bound in our model. 

Given the lower mass bound on $H^+$, it is useful to consider the implication for our model. 
Since we would like to consider the light $H$ with the mass below 10\,GeV, 
the dimensionless parameters in the scalar potential become large to obtain the ${\cal O}(100)$\,GeV mass difference.
The coupling $\lambda_3$, for instance, is about 2, when $m_{H^\pm} = 250$\,GeV and $m_H \leq {\cal O}(10)$\,GeV. 
This is consistent with the perturbativity, but 
one may worry that the Landau pole will appear at the low scale. 
We can resolve this problem by introducing one extra SM-singlet scalar, as discussed below.

\subsection{A simple extension without large quartic couplings}
\label{sec;extension}

In our setup, one additional scalar needs to be as light as DM to achieve the DM thermal production. 
As discussed above, however, the large quartic coupling in the scalar potential is required 
to obtain both light $H$ and heavy $H^{\pm}$. 
This large coupling is faced with the low scale Landau pole.
Here, we propose a simple extension to resolve the Landau-pole problem while realizing the light neutral scalar.

In order to avoid such a large coupling,
we can extend the model by introducing a real singlet scalar $S$. 
The mass of $S$ is set around the DM mass, while $H$ can be as heavy as $H^\pm$.
The Yukawa coupling between $S$ and $\psi$ is induced via the mass mixing between $S$
and $H$, so that $S$ is required to be $Z_2$-odd. 
In this case, the following term,
\beq 
- \Delta {\cal L} = A_S \, \Phi^\dagger \Phi_\nu S + h.c.
\eeq
is allowed in the Lagrangian. 
When $\Phi$ obtains the non-vanishing VEV, 
the mass mixing between the $H$ and $S$ is generated. 
The mass matrix is given by 
\beq
- {\cal L} \supset \frac{1}{2} 
\begin{pmatrix} H & S \end{pmatrix}
\begin{pmatrix} m_H^2 & \Delta m^2 \\ \Delta m^2 & m_S^2 \end{pmatrix} 
\begin{pmatrix} H \\ S \end{pmatrix} ,
\eeq
with $\Delta m^2 = {\rm Re} (A_S ) \, v$. 
The mass eigenstates are defined as 
\beq
\begin{pmatrix} H \\ S \end{pmatrix}
= 
\begin{pmatrix} \cos\theta & - \sin\theta \\ \sin\theta & \cos\theta \end{pmatrix}
\begin{pmatrix} h_2 \\ s \end{pmatrix} ,
\eeq
with the mixing angle being $\tan(2\theta) = 2\Delta m^2/ (m_H^2 - m_S^2)$.
The mass eigenvalues are expressed in terms of $m_H$, $m_S$ and $\theta$, 
\beq
m_{h_2}^2 = \frac{\cos^2\theta \, m_H^2 - \sin^2\theta \, m_S^2}{\cos(2\theta)} , \quad 
m_s^2 = \frac{\cos^2\theta \, m_S^2 - \sin^2\theta \, m_H^2}{\cos(2\theta)}.
\eeq
Therefore, with the small mixing, we get a light scalar by tuning $m_S$ as 
$m_S \simeq \tan\theta \, m_H$. 
Hereafter, we refer to the model with only one extra doublet as the minimal model, 
and the model presented here as the extended model. 
The extended model is similar to the model presented in Ref. \cite{Boehm:2013jpa} based on a supersymmetric (SUSY) extension of the SM. 
The prime difference is that DM is Dirac and a real scalar mediates the interaction with the neutrinos in our model, 
while the SUSY-like model involves Majorana DM (a neutralino) and a complex scalar mediator (a sneutrino). 
Such a difference causes a slight difference of the collider observables (e.g. the $Z$ boson decay), 
but both models are essentially the same. 
Hence, our results of the extended model update the previous work. 
Nonetheless, we would like to emphasize that our extended model has a unique DM signal. 
Since Dirac DM has the $s$-wave annihilation, 
our model can predict the significant neutrino flux from the galactic DM annihilation. 
This difference will help to discern the extended model from the SUSY-like model. 
The neutrino flux signals from DM will be studied in the next section. 

\begin{figure}[t]
\centering
\includegraphics[width=0.55\textwidth]{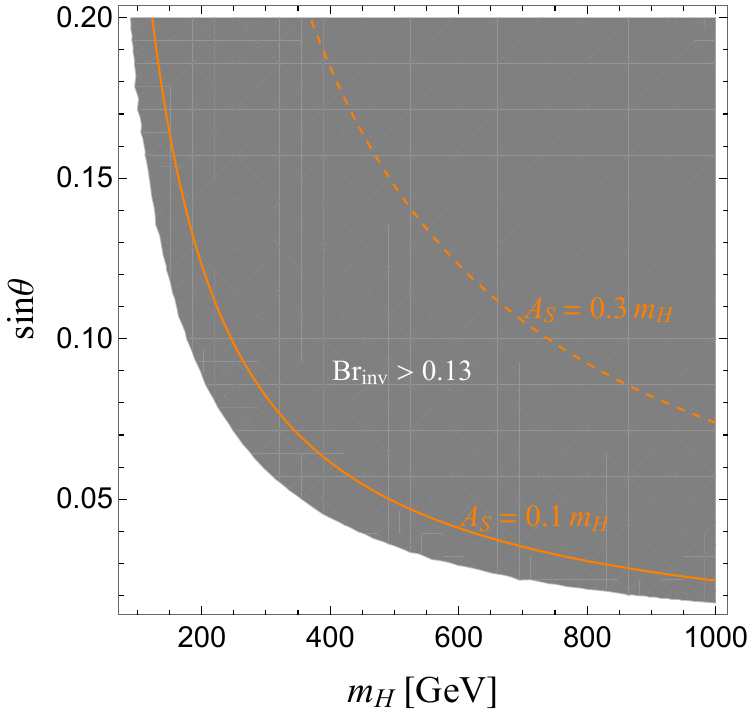}
\caption{The bound on the scalar mixing angle from the Higgs invisible decay in the extended model (gray). 
The orange lines represent the predictions of the mixing angle with 
$A_S = 0.1 \, m_H$ (solid) and $0.3 \,m_H$ (dashed).}
\label{fig;extension}
\end{figure}

Let us see how heavy the inert-doublet-like scalar can be. 
With the small mixing, the heavier scalar mass is expressed as 
$m_{h_2}^2 \simeq m_H^2 \simeq A_Sv/\theta$. 
The scalar mass depends on the ratio $A_S/m_H$ and mixing angle $\theta$. 
If we impose $A_S \leq 0.1 \, m_H$ for the vacuum stability, we find the upper limit as
$m_H \lesssim 0.1 \, v/\theta$. 
With $\theta=0.1$, for example, the mass should be less than $250$\,GeV (see also Fig.~\ref{fig;extension}). 
We can get a heavier $H$ for a smaller mixing. 

The size of the mixing angle is restricted by the Higgs invisible decay, 
since the $A$-term induces the new decay channel $h \to ss $ via the mixing. 
The $h$-$s$-$s$ coupling is given by 
\beq
- \Delta {\cal L} \supset - A_S \cos\theta \sin\theta \, h s^2 
\simeq - \frac{m_H^2}{v} \cos\theta \sin^2\theta \, h s^2 ,
\eeq
where $m_s \ll m_H$ is assumed. 
The partial decay width is given by 
\beq
\Gamma_{h\to ss} = \frac{\cos^2\theta \sin^4\theta \, m_H^4}{8\pi m_h v^2} \sqrt{1-\frac{4m_s^2}{m_h^2}} .
\end{equation}
The bound is shown in Fig.~\ref{fig;extension} with gray. 
It follows from the figure that the mixing angle will be at most 0.05 in light of the collider bounds. 
Such a small mixing, however, is enough to 
achieve the DM relic density in the thermal scenario. 
The required yukawa coupling $y^i_\nu$ is still small, ranging between 0.1 and 1, 
as far as the mass of $s$ is around the DM mass.

%===================
%===================

We comment on the direct search bound on the extended model from the LHC experiments.
The signals at the LHC are similar to those of the minimal model, except for the lightest scalar being $s$. 
The prime difference originates from the presence of the $A$-term. 
Since the heavy neutral scalars, $h_2$ and $A$, can decay to $s$ and $h$ via $A_S$, 
the $H^\pm h_2 \, (H^\pm A)$ productions followed by $h_2 \, (A) \to h s$ and $H^\pm \to W^\pm s$ 
produce the new signal, $W^\pm h$ plus the missing energy. 
This signal is studied in Ref. \cite{Aad:2019vvf} motivated by the supersymmetric model. 
Based on the analysis in Ref. \cite{Aad:2019vvf}, 
we can draw the bound on the charged scalar, but it is not so strong as $m_{H^\pm}\gtrsim 250$\,GeV
even with ${\rm Br}(h_2 \to h s)={\rm Br}(H^\pm \to W^\pm s)=1$. 
In the extended model, the charged scalar can further decay to $\ell^\pm \psi$ or $W^\pm h_2 \, (A)$ 
depending on the mass splitting of the scalars and the yukawa couplings $y_\nu^i$, 
so that the actual signal is a bit complicated. 
The detailed analysis is beyond the scope of this paper, but as an existence proof, 
we shall briefly show in the following that the extended model is experimentally viable in fact. 

As a benchmark point, we consider the scalar mass spectrum $m_A = m_{H^+} = m_{h_2} = 300\,{\rm GeV}$ and the scalar mixing angle $\sin\theta=0.05$, that we will take later in the DM phenomenology. 
We also assume DM couples only to the third generation lepton with the yukawa coupling $y_\nu^\tau=0.16$, 
which can explain the observed DM abundance and avoid all limits from DM search experiments as shown in Sec.~\ref{sec4}. 
In $\sqrt{s}=13$\,TeV $pp$ collisions, 
the pair-production cross section of the 300\,GeV $H^{\pm}$ 
is estimated as 3.15\,fb using {\tt MadGraph5\_aMC@NLO} \cite{Alwall:2014hca}. 
For the benchmark point, 
the branching ratio of $H^\pm \to \tau^\pm \psi$ is about 0.9\footnote{In the extended model, 
the decay width of $H^\pm$ into $W^\pm s$ is suppressed by the small scalar mixing $\sin\theta \ll 1$, 
so that the branching fraction ${\rm Br}(H^\pm \to \ell^\pm \psi)$ can exceed 1/2. 
This is a difference from the minimal model.}, 
which means $\sigma(pp \to H^+ H^-) \cdot {\rm Br}(H^\pm \to \tau^\pm\psi)^2 \simeq 2.55$\,fb. 
On the other hand, the ATLAS stau search~\cite{Aad:2019byo} indicates that 
the lower limit on the charged scalar mass is 310\,GeV 
with ${\rm Br}(H^\pm \to \tau^\pm \psi)=1$ and $m_\psi \ll m_{H^+}$, 
for which we estimate $\sigma(pp \to H^+ H^-) \simeq 2.70$\,fb. 
Hence, the benchmark point is expected to evade the ATLAS bound, 
assuming the acceptance is the same as the stau search. 
There is another process, $pp \to H^+ H^- \to W^+ W^- ss$, 
which produces a signal of dilepton plus large missing energy. 
However, it provides much weak constraint due to the small branching fraction ${\rm Br}(H^\pm \to W^\pm s) \simeq 0.1$.
We note that the pair-productions of the other scalars do not provide any stringent bound, 
although their cross sections are comparable with that of $H^+H^-$: 2.95 fb ($h_2A$), 5.54 fb, ($H^\pm h_2$), and 5.54 fb ($H^\pm A$). 
This is mainly because the neutral scalars mostly decay to the invisible particles, $\psi$ and $\nu$. 
Then, the main signal from these productions is mono-lepton plus missing energy, that is not strongly constrained. 
Therefore, the benchmark point can escape the current collider bounds 
and is an experimentally viable parameter set. 

In general, we can consider the heavier scalars, 
for which the production cross section significantly drops. 
Finite mass splittings between the scalars will also have some impacts on actual signals and 
may weaken collider bounds. 
Hence, the collider search does not provide a serious constraint yet. 
Moreover, it is easy to avoid the other constraints on the extra scalars in the extended model. 
Since the quartic couplings in the scalar potential are released 
from the strong relation $\la_3 \approx - (\la_4+\la_5)$, 
we can take a small $\la_3$ while keeping a heavy $H^+$. 
This allows a little deviation of the Higgs signal strength $R_{\gamma\gamma}$. 
The precision observables, such as the oblique corrections and the $Z$-pole observables, 
bring no severe limits, as far as $h_2 \,(\approx H)$, $A$ and $H^\pm$ are around the EW scale 
and the mass splittings are not large. 
Therefore, we conclude that 
the extended model proposed in this subsection still has a large allowed parameter space. 
The detailed study in the collider search for the extra scalars will be followed up in a future work.

\section{DM physics}
\label{sec4}

Now that we have the light scalar in the minimal (or extended) lepton portal model, 
let us examine the DM phenomenology in the light mass region. 
The relevant interaction for DM physics is in the form of 
\begin{equation}
- {\cal L}_\ell = y_\nu^i \left[ \frac{1}{\sqrt{2}} \overline{\nu^i_L} (H-iA) \psi_R 
- \overline{e^i_L} H^- \psi_R \right] + h.c. ,
\label{eq;yukawa}
\end{equation}
where $H$ is the lightest neutral scalar in the minimal model, 
or $H = \cos\theta \, h_2 - \sin\theta \, s$ in the extended model. 
We will consider two cases where only the DM and one of the neutral scalars are light, 
while the others are heavy: 
\begin{align}
{\rm (I)}~& \mbox{minimal model:} ~~ m_\psi < m_H \ll m_A = m_{H^+} = 300\,{\rm GeV}, \\
{\rm (II)}~& \mbox{extended model:} ~~ m_\psi < m_s \ll m_A = m_{H^+} = m_{h_2} = 300\,{\rm GeV} 
\mbox{   and   } \sin\theta=0.05
\end{align}
If $m_\psi > m_H$ or $m_s$, the scalar can be another DM candidate, but we do not study this case. 
In the following discussion, 
we assume that the coupling only to the third generation leptons is non-vanishing, 
mainly to avoid the collider bounds and large charged lepton flavor violations. 
Then, we have only three parameters, $m_\psi$, $m_H$ (or $m_s$) and $y_\nu$, 
where $y_\nu$ denotes the Yukawa coupling with the third-generation leptons.

We would like to mention the similarity and distinction between the minimal and the extended models in the DM physics. 
As seen in Eq.(\ref{eq;yukawa}), 
the coupling of DM to the charged scalar is written by $y_\nu$ in both the minimal and extended models, 
so that the DM physics that the charged scalar is involved in is completely the same. 
Indeed, as we will show in Sec.~\ref{sec;constraints}, 
experimental constraints on DM induced by this coupling are common to both models. 
On the other hand, the coupling of DM to the lightest neutral scalar, which is mostly responsible for the DM production, is $y_\nu$ in the minimal model, while it is $y_\nu \sin\theta$ in the extended model. 
Thus, the DM physics induced by the neutral scalar exchanging, {\it i.e.} DM production, differs each other. 
Nevertheless, there is an aesthetic correspondence between these two models, 
and once we make the replacement $y_\nu \to y_\nu \sin\theta$ and $m_H \to m_s$ in the minimal model, 
the DM phenomenology in the minimal model can be readily converted into the one in the extended model.

\subsection{Relic density}

We assume the thermal freeze-out mechanism for DM production. 
The main annihilation mode is the one into neutrinos via the $t$-channel neutral scalar exchanging.  
The cross sections are given, in the minimal model, by 
\begin{align}
(\sigma v_{\rm rel})_{\psi\bar{\psi}\to\nu\bar{\nu}} & 
\simeq \frac{y_\nu^4 m_\psi^2}{128\pi (m_\psi^2+m_H^2-m_\nu^2)^2} \sqrt{1-\frac{m_\nu^2}{m_\psi^2}}, \\
(\sigma v_{\rm rel})_{\psi\psi\to\nu \nu} & = (\sigma v)_{\bar{\psi}\bar{\psi}\to\bar{\nu}\bar{\nu}} 
\simeq \frac{y_\nu^4 (2m_\psi^2+m_\nu^2)}{256\pi (m_\psi^2+m_H^2-m_\nu^2)^2} \sqrt{1-\frac{m_\nu^2}{m_\psi^2}}, 
\end{align}
where $v_{\rm rel}$ denotes the relative velocity of DM and we only keep the partial $s$-wave. 
The cross sections in the extended model are obtained by the replacement, 
$m_H \to m_s$ and $y_\nu \to \sin\theta \, y_\nu$. 
We neglect the sub-leading contributions from the heavier particles in the above equations, but 
we take into account all possible contributions numerically in our relic density calculation by employing {\tt micrOMEGAs\_4\_3\_5}~\cite{Belanger:2014vza}. 

\begin{figure}[t]
\includegraphics[width=0.5\textwidth]{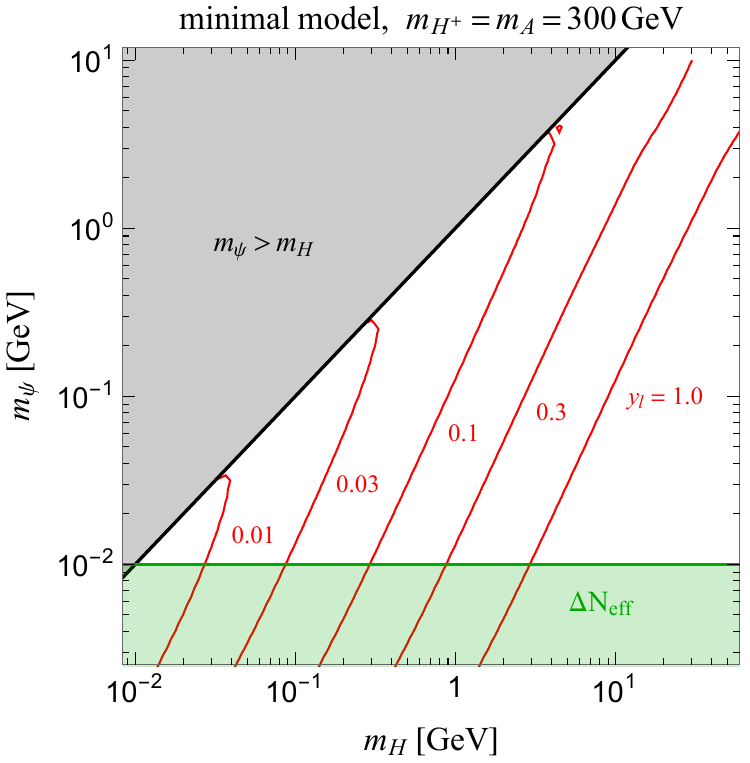}
\includegraphics[width=0.5\textwidth]{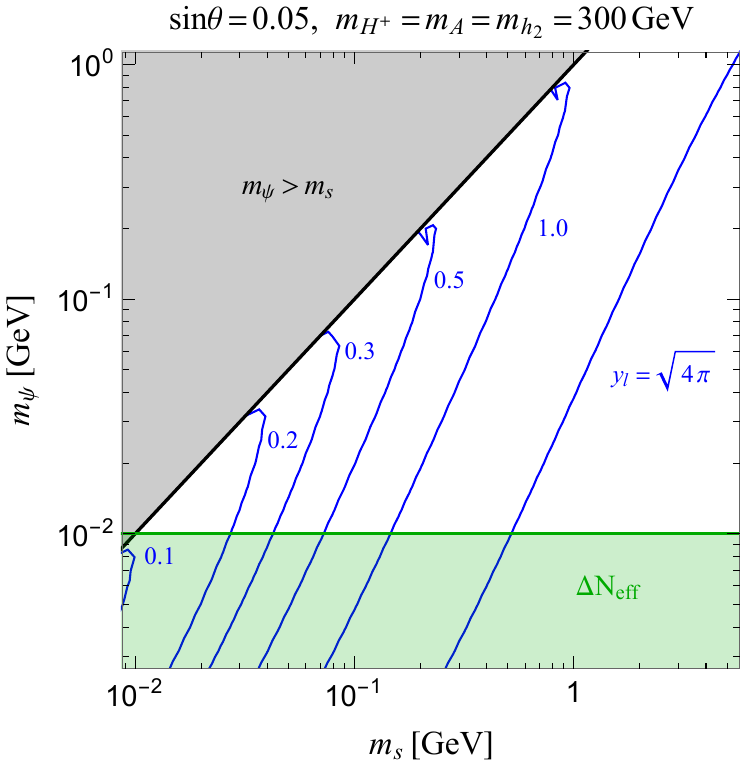}
\caption{Allowed region and the size of the yukawa coupling to explain the DM relic density. 
(Left) the minimal model with $m_{H^+} = m_A = 300$\,GeV, 
(right) the extended model with $m_{H^+} = m_A = m_{h_2} = 300$\,GeV and $\sin\theta=0.05$.}
\label{fig;allowed}
\end{figure}

In Fig.~\ref{fig;allowed}, we show the allowed region and 
the size of the Yukawa coupling that is required for the thermal production. 
The left panel corresponds to the minimal model and the right panel to the extended model. 
In both cases, the DM mass below 10\,MeV is excluded, since such a light DM increases the effective number of neutrinos $N_{\rm eff}$~\cite{Boehm:2013jpa,Nollett:2014lwa,Heo:2015kra,Sabti:2019mhn}. 
In the minimal model, DM is consistently produced in a wide mass range. 
Even relatively high mass over 10\,GeV is possible, while keeping the perturbativity $y_\nu\leq1$. 
The Yukawa coupling takes the value of ${\cal O}$(0.01--1), 
depending on the mass splitting of DM and $H$. 
In the extended model, the small scalar mixing suppresses the annihilation cross section, 
so that the larger Yukawa coupling is required. 
As a result, the DM mass is only allowed up to 800\,MeV, 
if we impose the perturbativity condition, $y_\nu\leq1$.

%===================
%===================

%===================
\begin{figure}[t]
\includegraphics[width=0.5\textwidth]{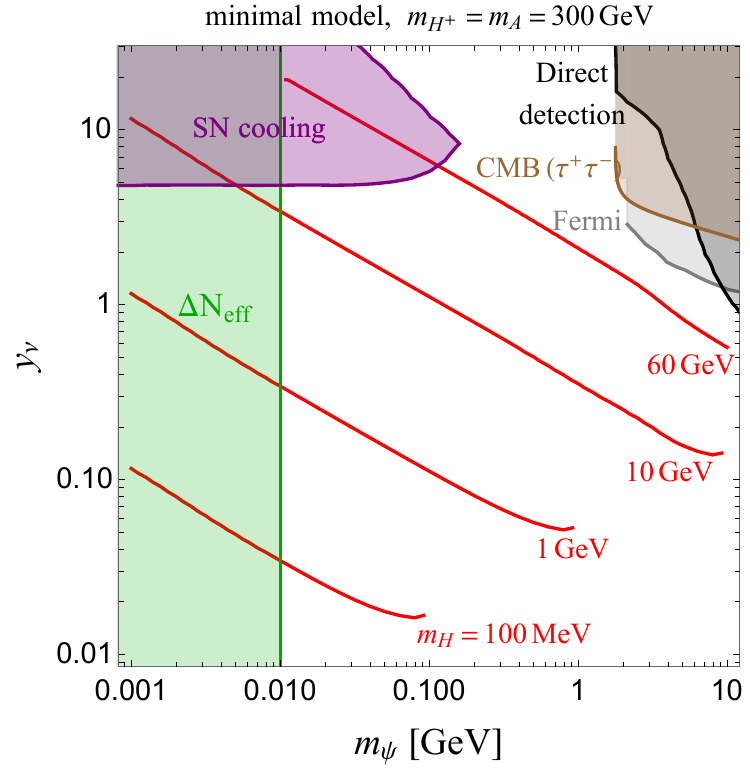}
\includegraphics[width=0.5\textwidth]{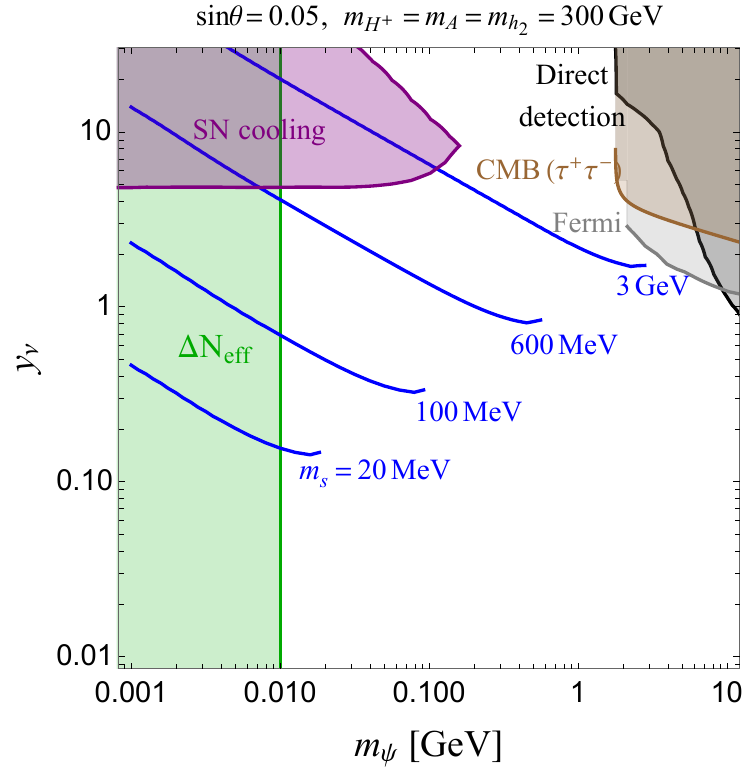}
\caption{Experimental upper limits on the yukawa coupling $y_\nu$ in the minimal model (left) and extended model (right). The shaded regions are excluded by direct and indirect DM search experiments. 
The red and blue lines show the values of the lightest neutral scalar mass that are required to explain the DM abundance.}
\label{fig;constraints}
\end{figure}
%===================

\subsection{Constraints from the DM experiments}
\label{sec;constraints}

We next discuss the experimental constraints concerned with DM in the light mass window. 
The Yukawa coupling $y_\nu$ relevant to the DM production induces 
the DM annihilation into $\tau$ leptons at tree level via the $t$-channel charged scalar exchanging. 
This annihilation is not responsible for the DM production, 
but it may be confronted with the CMB data and 
the galactic gamma-ray searches in the light mass region.
The same coupling further generates the electromagnetic (EM) form factors of DM 
via the one-loop diagram involving the charged scalar~\cite{Ibarra:2015fqa,Kawamura:2020qxo}. 
The induced form factors allow DM to scatter with the ordinary matter, 
that is strongly restricted by the direct detection experiments~\cite{Chu:2018qrm}.  
In the following, we will make sure that 
such constraints do not conflict with the DM thermal production. 

In Fig.~\ref{fig;constraints}, 
we illustrate the leading constraints from the direct detection and indirect searches 
in the ($m_\psi, \, y_\nu$) plane. 
The left panel corresponds to the minimal model and the right panel to the extended model. 
The red and blue lines represent the values of the light scalar mass 
that are required to fit the DM relic abundance. 

The direct detection gives the best sensitivity to the Yukawa coupling $y_\nu$ 
in the mass region above $10$\,GeV. 
The strongest bound (black) is set by the measurements of the DM-nucleon scattering at 
XENON1T ($5\,{\rm GeV} \lesssim m_\psi$) \cite{Aprile:2018dbl}, 
XENON1T with ionization signals ($3\,{\rm GeV} \lesssim m_\psi \lesssim 5\,{\rm GeV}$) \cite{Aprile:2019xxb}, 
DarkSide50 ($2\,{\rm GeV} \lesssim m_\psi \lesssim 3\,{\rm GeV}$) \cite{Agnes:2018ves}, 
XENON1T with migdal effects ($0.1\,{\rm GeV} \lesssim m_\psi \lesssim 2\,{\rm GeV}$) \cite{Aprile:2019jmx} 
and a {\tt TEA-LAB} simulated experiment inspired by the DarkSide50 result 
($0.05\,{\rm GeV} \lesssim m_\psi \lesssim 0.1\,{\rm GeV}$) \cite{GrillidiCortona:2020owp}. 
The scattering with the electron is also induced by photon exchanging. 
While several experiments have recently been improving the sensitivity to the DM-electron scattering in the sub-GeV region~\cite{Aprile:2019xxb,Agnes:2018oej,Essig:2012yx,Barak:2020fql},  
the limits are still weak and we do not find them in the figure.

In the mass region below 10\,GeV, the DM annihilation into 2$\tau$ provides the best limit, 
that is set by 11-year Fermi-LAT observations of the diffuse gamma-rays from dwarf spheroidal galaxies~\cite{Hoof:2018hyn}. 
It covers the DM mass above $m_\tau$. 
The same annihilation mode also contributes to an extra ionization in the post-recombination era, 
modifying the anisotropies of the CMB and in turn limiting the annihilation rate. 
The resulting upper limit on the DM annihilation cross section is evaluated as 
$(\sigma v)_{\tau\bar{\tau}} /(2m_\psi) \lesssim 5.1 \times 10^{-27}\,{\rm cm^3/s/GeV}$\cite{Slatyer:2015jla,Leane:2018kjk}. 
Another limit comes from an observation of the supernova (SN) cooling (purple). 
The DM particles with the light mass can be created inside the SN thought the EM form factors. 
If the created DM particles escape from the core, it increases the SN cooling rate. 
One can set the SN cooling limit when the increased cooling rate significantly deviates from 
the prediction in the standard cooling theory~\cite{Raffelt:1996wa}. 
We evaluate the limit on the Yukawa coupling by translating the SN cooling limit on the EM form factors of DM~\cite{Chu:2018qrm}. 

Note that all the experimental constraints above, except for the $\Delta N_{\rm eff}$ bound, 
arise from the DM coupling to the charged scalar and thus depend on the charged scalar mass $m_{H^+}$. 
However, its dependence is very simple. 
As the charged scalar mass becomes twice, the constraints on the yukawa coupling $y_\nu$ become weaker uniformly by a factor of 2\footnote{Strictly speaking, the bounds slightly depend on the DM mass as well. However, its contribution is suppressed by $m_\psi^2/m_{H^+}^2$, compared with the leading contribution, and hence very small. We include the sub-leading contributions to draw the plots in Fig.~\ref{fig;constraints}.}. 
Since the collider bound is found to be $m_{H^+} \geq 250$\,GeV in our models, 
the limits shown in Fig.~\ref{fig;constraints} are almost the strongest ones. 
Therefore, we conclude that the experimental constraints can easily be evaded in our models, 
as far as we consider the DM mass below 10\,GeV and the perturbative coupling $y_\nu \leq 1$.

\subsection{Searching for light mass region with neutrino telescopes}

We have found a new parameter space in the lepton portal DM model 
where DM has a light mass of 10\,MeV--10\,GeV. 
In this region, the model is almost free from the experimental constraints. 
Even the future planned experiments of direct detection and gamma-ray searches 
will not have the ability to probe it. 
Therefore, it is natural to explore the way to test such a light mass region. 
In this section, we suggest that neutrino flux observations with neutrino telescopes 
will be able to indirectly detect the light mass DM in our model. 

The expected electron neutrino flux originated from DM annihilation in our galaxy is expressed by 
\begin{equation}
\frac{d\Phi_{\nu_e}}{dE_\nu} 
= \frac{1}{4\pi} \sum_i \frac{\VEV{\sigma v}_i}{4m_\psi^2} \kappa \frac{dN_i}{dE_\nu} 
\int dx d\Omega \, \rho_{\rm DM}(r(x,\Omega))^2,
\label{eq:flux}
\end{equation}
where 
$\VEV{\sigma v}_i$ denotes the annihilation cross section into a final state $i$, 
$dN_i/dE_\nu$ the neutrino spectral function for the final state $i$, 
$\kappa$ a model dependent constant which characterizes electron-neutrino flavor fraction, 
and $\rho(r)$ the DM density profile in the galactic halo. 
The line-of-sight integral of the DM density square 
is called the astrophysical $J$-factor. 
We consider a Navarro-Frenk-White (NFW) profile~\cite{Navarro:1996gj} 
with parameters given in \cite{Vertongen:2011mu}, 
that leads to the all-sky $J$-factor, $J \simeq 1.58 \times 10^{23}\,{\rm GeV^2/cm^5}$. 
Since DM annihilates into a neutrino pair in our model, 
the neutrino spectrum takes the monochromatic form, $dN/dE_\nu \propto \delta(E_\nu-m_\psi)$. 

The extra neutrino flux produced from the galactic DM annihilation travels to neutrino telescopes 
and scatters off nuclei in the detector to produce DM signals. 
The signal events are compared with the results measured at the SK, KamLAND, Borexino detectors 
to derive upper limits on the annihilation cross section in \cite{Yuksel:2007ac,PalomaresRuiz:2007eu,Primulando:2017kxf,Campo:2017nwh,Klop:2018ltd,Arguelles:2019ouk}. 
They also show the projected sensitivity of the future neutrino telescopes, such as HK, DUNE and JUNO~\cite{Campo:2018dfh,Klop:2018ltd,Arguelles:2019ouk,Bell:2020rkw,Asai:2020qlp}. 
We do not perform new analysis of the DM signal in this paper, 
but refer to these upper limits and demonstrate the capability of testing the light DM in our model. 

\begin{figure}[t]
\centering
\includegraphics[width=0.7\textwidth]{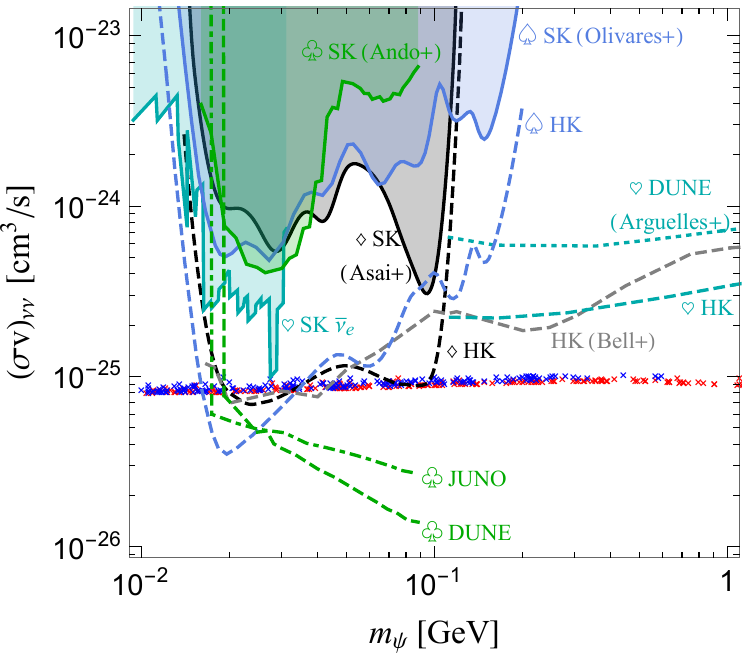}
\caption{Upper limits on the DM annihilation cross section into neutrinos. 
The shaded regions are excluded by the current experimental results, while 
the dashed, dotted and dot-dashed lines stand for the future sensitivity reach. 
The lines with the same mark ($\spadesuit$, $\clubsuit$, $\heartsuit$, $\diamondsuit$) are extracted from the same paper: 
$\spadesuit$ Olivares et al. \cite{Campo:2017nwh,Campo:2018dfh}, 
$\clubsuit$ Ando et al.  \cite{Klop:2018ltd}, $\heartsuit$ Arguelles et al. \cite{Arguelles:2019ouk}, $\diamondsuit$ Asai et al. \cite{Asai:2020qlp} and Bell et al. \cite{Bell:2020rkw}.
The ``$\times$'' points represent the predictions 
in the minimal model (red) and extended model (blue).}
\label{fig;indirect}
\end{figure}

In Fig.~\ref{fig;indirect}, 
we summarize the current upper limits on and the future sensitivity to the annihilation cross section into neutrinos. 
It is assumed the flavor ratio at detector is $\nu_e : \nu_\mu : \nu_\tau = 1 : 1 : 1$ that 
leads to $\kappa=1/3$ in Eq.(\ref{eq:flux}). 
The limits with the same mark ($\spadesuit$, $\clubsuit$, $\heartsuit$, $\diamondsuit$) are extracted from the same paper: 
$\spadesuit$ Olivares et al. \cite{Campo:2017nwh,Campo:2018dfh}, 
$\clubsuit$ Ando et al.  \cite{Klop:2018ltd}, $\heartsuit$ Arg\"uelles et al. \cite{Arguelles:2019ouk}, $\diamondsuit$ Asai et al. \cite{Asai:2020qlp} and Bell et al. \cite{Bell:2020rkw}. 
In referring to these limits, we appropriately adjust the neutrino flavor fraction $\kappa$. 
The shaded region is excluded by the current observations, 
while the dash, dotted and dot-dashed lines represent the future sensitivity. 
We also plot the predictions in the minimal and extended models with the red and blue ``$\times$'' points, respectively. 
When plotting, we scan the allowed parameter space in Fig.~\ref{fig;allowed}, 
where DM mass is heavier than 10\,MeV and the Yukawa coupling is smaller than unity. 
In our model, the vertical axis is interpreted as  
$(\sigma v)_{\nu\nu} \equiv (\sigma v)_{\psi\bar{\psi} \to \nu\bar{\nu}} 
+ \frac{1}{2} \left\{ (\sigma v)_{\psi\psi \to \nu\nu} + (\sigma v)_{\bar{\psi}\bar{\psi} \to \bar{\nu}\bar{\nu}}\right\}$. 

%===================
%===================

We see in the figure that the model mostly predicts $(\sigma v)_{\nu\nu} \simeq 10^{-25}\,{\rm cm^3/s}$,  
that corresponds to the canonical thermal relic cross section for the sub-GeV Dirac DM~\cite{Steigman:2012nb,Saikawa:2020swg}. 
The predictions are distributed equally in the minimal and extended setups. 
The current best limit is set by 
a low-energy neutrino flux observation at the SK-IV \cite{Arguelles:2019ouk} ($m_\psi \lesssim 30$\,MeV) and 
results of the supernova relic neutrino (SRN) searches at the SK \cite{Campo:2017nwh,Campo:2018dfh,Klop:2018ltd,Arguelles:2019ouk,Bell:2020rkw,Asai:2020qlp} ($30\,{\rm MeV}\lesssim m_\psi \lesssim 200\,{\rm MeV}$). 
None of the current experiments can still cover the thermal relic cross section. 
However, the sensitivity of the future HK experiment will be able to test this DM candidate with the 20--30\,MeV mass. 
The DUNE and JUNO experiments will have better sensitivity and 
cover the mass range from 20\,MeV to 100\,MeV. 
In the higher mass region ($100\,{\rm MeV} \leq m_\psi$), unfortunately, 
even the future telescopes do not reach the model predictions by a factor of 3. 
We hope that further improvements or developments in future will grow the sensitivity by a factor of several 
and will be able to cover the wide parameter space. 

We would like to add that collider searches can provide complementary bounds to 
the neutrino flux searches. If future collider experiments can search for the charge Higgs $H^+$ 
with a heavier mass and find no affirmative signal, the minimal model is driven into a corner, 
since it requires an extremely large quartic coupling, that breaks perturbativity, in that case. 
In the extended model, the improvement of the bound on the charged scalar mass will indirectly put an upper limit on the DM mass. 
As the charged scalar is heavier, a smaller scalar mixing angle is required due to the Higgs invisible decay bound. 
Then, the DM thermal production requires a larger Yukawa coupling and, in turn, 
a light mediator and light DM. 
The current upper limit on the DM mass is about 800\,MeV, but if it is improved to be 100\,MeV, 
we will be able to test the most parameter space by combining with future neutrino flux observations.

\section{Summary}
\label{sec5}

We have explored a novel possibility that the lepton portal DM has a light mass below 10\,GeV. 
The DM abundance is thermally produced by its annihilation into neutrinos. 
It is essential to introduce a light scalar mediator for the successful DM production. 
In the minimal setup, where only one extra doublet scalar is added,
the light neutral scalar is realized by adjusting the quartic couplings in the scalar potential. 
In this case, the constraint from the direct search for the charged scalar requires a very large mass splitting 
between the charged and the neutral scalars, that corresponds to the large quartic coupling.
Then, we have found that the minimal setup predicts explicit deviations of the observables relevant to
the 125 GeV Higgs boson $h$.  The partial decay width of $h$ into 2$\gamma$ becomes 10\% smaller than the SM prediction.
The branching ratio of the Higgs invisible decay is estimated as ${\cal O}(0.1)$, that can be tested near future.

If we introduce a SM singlet scalar in addition to the EW doublet scalar, the quartic couplings in the scalar potential
can be small. Even in this extended model, a light neutral scalar is required to obtain the correct DM abundance thermally. 
Figure \ref{fig;extension} shows that the constraint from the invisible Higgs decay limits the heavier neutral scalar mass and the mass mixing between the two neutral scalars.

Based on the constraints concerned with the EWPOs and the 125 GeV Higgs, 
we have studied the DM physics in both minimal and extended models. 
We have shown that DM is correctly produced in the wide mass range, 
without any conflict with the current experimental bounds and the cosmological observations. 
See Figs.~\ref{fig;allowed} and \ref{fig;constraints} for the allowed parameter region in each model. 
Moreover, we have demonstrated that 
the allowed region predicts the significant neutrino flux from DM annihilation in our galaxy, 
that could be within the experimental reach of the future neutrino telescopes, 
such as HK, DUNE and JUNO. 
We expect that our models will be fully tested by combining the neutrino flux search 
with results of search for the new EW charged scalars at the LHC and the future collider experiments.

\section*{Acknowledgement}

SO would like to thank the hospitality of the Theoretical Particle Physics Group in Kindai University, where this work was initiated during his visit. 
This work is supported in part by NSERC of Canada (SO) and the Grant-in-Aid for Scientific Research from the
Ministry of Education, Science, Sports and Culture (MEXT), 
Japan  No. 19H04614, No. 19H05101 and No. 19K03867 (YO).

{\small
\bibliographystyle{JHEP}
\bibliography{ref_lightDM}

\providecommand{\href}[2]{#2}\begingroup\raggedright\begin{thebibliography}{10}

\bibitem{Aprile:2018dbl}
{\bf XENON} Collaboration, E.~Aprile et~al., {\it {Dark Matter Search Results
  from a One Ton-Year Exposure of XENON1T}},  {\em Phys. Rev. Lett.} {\bf 121}
  (2018), no.~11 111302, [\href{http://arxiv.org/abs/1805.12562}{{\tt
  arXiv:1805.12562}}].

\bibitem{Slatyer:2015jla}
T.~R. Slatyer, {\it {Indirect dark matter signatures in the cosmic dark ages.
  I. Generalizing the bound on s-wave dark matter annihilation from Planck
  results}},  {\em Phys. Rev. D} {\bf 93} (2016), no.~2 023527,
  [\href{http://arxiv.org/abs/1506.03811}{{\tt arXiv:1506.03811}}].

\bibitem{Leane:2018kjk}
R.~K. Leane, T.~R. Slatyer, J.~F. Beacom, and K.~C. Ng, {\it {GeV-scale thermal
  WIMPs: Not even slightly ruled out}},  {\em Phys. Rev. D} {\bf 98} (2018),
  no.~2 023016, [\href{http://arxiv.org/abs/1805.10305}{{\tt
  arXiv:1805.10305}}].

\bibitem{Hoof:2018hyn}
S.~Hoof, A.~Geringer-Sameth, and R.~Trotta, {\it {A Global Analysis of Dark
  Matter Signals from 27 Dwarf Spheroidal Galaxies using 11 Years of Fermi-LAT
  Observations}},  {\em JCAP} {\bf 02} (2020) 012,
  [\href{http://arxiv.org/abs/1812.06986}{{\tt arXiv:1812.06986}}].

\bibitem{Primulando:2017kxf}
R.~Primulando and P.~Uttayarat, {\it {Dark Matter-Neutrino Interaction in Light
  of Collider and Neutrino Telescope Data}},  {\em JHEP} {\bf 06} (2018) 026,
  [\href{http://arxiv.org/abs/1710.08567}{{\tt arXiv:1710.08567}}].

\bibitem{Campo:2017nwh}
A.~Olivares-Del~Campo, C.~B\oe~hm, S.~Palomares-Ruiz, and S.~Pascoli, {\it
  {Dark matter-neutrino interactions through the lens of their cosmological
  implications}},  {\em Phys. Rev. D} {\bf 97} (2018), no.~7 075039,
  [\href{http://arxiv.org/abs/1711.05283}{{\tt arXiv:1711.05283}}].

\bibitem{Foldenauer:2018zrz}
P.~Foldenauer, {\it {Light dark matter in a gauged $U(1)_{L_\mu-L_\tau}$
  model}},  {\em Phys. Rev. D} {\bf 99} (2019), no.~3 035007,
  [\href{http://arxiv.org/abs/1808.03647}{{\tt arXiv:1808.03647}}].

\bibitem{Bernreuther:2020koj}
E.~Bernreuther, S.~Heeba, and F.~Kahlhoefer, {\it {Resonant Sub-GeV Dirac Dark
  Matter}},  \href{http://arxiv.org/abs/2010.14522}{{\tt arXiv:2010.14522}}.

\bibitem{Asai:2020qlp}
K.~Asai, S.~Okawa, and K.~Tsumura, {\it {Search for U(1)$_{L_\mu-L_\tau}$
  charged Dark Matter with neutrino telescope}},
  \href{http://arxiv.org/abs/2011.03165}{{\tt arXiv:2011.03165}}.

\bibitem{Batell:2017cmf}
B.~Batell, T.~Han, D.~McKeen, and B.~Shams Es~Haghi, {\it {Thermal Dark Matter
  Through the Dirac Neutrino Portal}},  {\em Phys. Rev. D} {\bf 97} (2018),
  no.~7 075016, [\href{http://arxiv.org/abs/1709.07001}{{\tt
  arXiv:1709.07001}}].

\bibitem{McKeen:2018pbb}
D.~McKeen and N.~Raj, {\it {Monochromatic dark neutrinos and boosted dark
  matter in noble liquid direct detection}},  {\em Phys. Rev. D} {\bf 99}
  (2019), no.~10 103003, [\href{http://arxiv.org/abs/1812.05102}{{\tt
  arXiv:1812.05102}}].

\bibitem{Blennow:2019fhy}
M.~Blennow, E.~Fernandez-Martinez, A.~Olivares-Del~Campo, S.~Pascoli,
  S.~Rosauro-Alcaraz, and A.~Titov, {\it {Neutrino Portals to Dark Matter}},
  {\em Eur. Phys. J. C} {\bf 79} (2019), no.~7 555,
  [\href{http://arxiv.org/abs/1903.00006}{{\tt arXiv:1903.00006}}].

\bibitem{Bai:2014osa}
Y.~Bai and J.~Berger, {\it {Lepton Portal Dark Matter}},  {\em JHEP} {\bf 08}
  (2014) 153, [\href{http://arxiv.org/abs/1402.6696}{{\tt arXiv:1402.6696}}].

\bibitem{Chang:2014tea}
S.~Chang, R.~Edezhath, J.~Hutchinson, and M.~Luty, {\it {Leptophilic Effective
  WIMPs}},  {\em Phys. Rev.} {\bf D90} (2014), no.~1 015011,
  [\href{http://arxiv.org/abs/1402.7358}{{\tt arXiv:1402.7358}}].

\bibitem{Kawamura:2020qxo}
J.~Kawamura, S.~Okawa, and Y.~Omura, {\it {Current status and muon $g-2$
  explanation of lepton portal dark matter}},  {\em JHEP} {\bf 08} (2020) 042,
  [\href{http://arxiv.org/abs/2002.12534}{{\tt arXiv:2002.12534}}].

\bibitem{Ibarra:2015fqa}
A.~Ibarra and S.~Wild, {\it {Dirac dark matter with a charged mediator: a
  comprehensive one-loop analysis of the direct detection phenomenology}},
  {\em JCAP} {\bf 05} (2015) 047, [\href{http://arxiv.org/abs/1503.03382}{{\tt
  arXiv:1503.03382}}].

\bibitem{Steigman:2012nb}
G.~Steigman, B.~Dasgupta, and J.~F. Beacom, {\it {Precise Relic WIMP Abundance
  and its Impact on Searches for Dark Matter Annihilation}},  {\em Phys. Rev.
  D} {\bf 86} (2012) 023506, [\href{http://arxiv.org/abs/1204.3622}{{\tt
  arXiv:1204.3622}}].

\bibitem{Saikawa:2020swg}
K.~Saikawa and S.~Shirai, {\it {Precise WIMP Dark Matter Abundance and Standard
  Model Thermodynamics}},  {\em JCAP} {\bf 08} (2020) 011,
  [\href{http://arxiv.org/abs/2005.03544}{{\tt arXiv:2005.03544}}].

\bibitem{Barbieri:2006dq}
R.~Barbieri, L.~J. Hall, and V.~S. Rychkov, {\it {Improved naturalness with a
  heavy Higgs: An Alternative road to LHC physics}},  {\em Phys. Rev. D} {\bf
  74} (2006) 015007, [\href{http://arxiv.org/abs/hep-ph/0603188}{{\tt
  hep-ph/0603188}}].

\bibitem{Belyaev:2016lok}
A.~Belyaev, G.~Cacciapaglia, I.~P. Ivanov, F.~Rojas-Abatte, and M.~Thomas, {\it
  {Anatomy of the Inert Two Higgs Doublet Model in the light of the LHC and
  non-LHC Dark Matter Searches}},  {\em Phys. Rev. D} {\bf 97} (2018), no.~3
  035011, [\href{http://arxiv.org/abs/1612.00511}{{\tt arXiv:1612.00511}}].

\bibitem{Belyaev:2018ext}
A.~Belyaev, T.~Fernandez Perez~Tomei, P.~Mercadante, C.~Moon, S.~Moretti,
  S.~Novaes, L.~Panizzi, F.~Rojas, and M.~Thomas, {\it {Advancing LHC probes of
  dark matter from the inert two-Higgs-doublet model with the monojet signal}},
   {\em Phys. Rev. D} {\bf 99} (2019), no.~1 015011,
  [\href{http://arxiv.org/abs/1809.00933}{{\tt arXiv:1809.00933}}].

\bibitem{Zyla:2020zbs}
{\bf Particle Data Group} Collaboration, P.~Zyla et~al., {\it {Review of
  Particle Physics}},  {\em PTEP} {\bf 2020} (2020), no.~8 083C01.

\bibitem{ATLAS:2020cjb}
{\bf ATLAS} Collaboration, {\it {Search for invisible Higgs boson decays with
  vector boson fusion signatures with the ATLAS detector using an integrated
  luminosity of 139 fb$^{-1}$}}, .

\bibitem{Sirunyan:2018owy}
{\bf CMS} Collaboration, A.~M. Sirunyan et~al., {\it {Search for invisible
  decays of a Higgs boson produced through vector boson fusion in proton-proton
  collisions at $\sqrt{s} =$ 13 TeV}},  {\em Phys. Lett. B} {\bf 793} (2019)
  520--551, [\href{http://arxiv.org/abs/1809.05937}{{\tt arXiv:1809.05937}}].

\bibitem{deBlas:2019rxi}
J.~de~Blas et~al., {\it {Higgs Boson Studies at Future Particle Colliders}},
  {\em JHEP} {\bf 01} (2020) 139, [\href{http://arxiv.org/abs/1905.03764}{{\tt
  arXiv:1905.03764}}].

\bibitem{Kanemura:2016sos}
S.~Kanemura, M.~Kikuchi, and K.~Sakurai, {\it {Testing the dark matter scenario
  in the inert doublet model by future precision measurements of the Higgs
  boson couplings}},  {\em Phys. Rev. D} {\bf 94} (2016), no.~11 115011,
  [\href{http://arxiv.org/abs/1605.08520}{{\tt arXiv:1605.08520}}].

\bibitem{ATLAS:2019jst}
{\bf ATLAS} Collaboration, {\it {Measurements and interpretations of
  Higgs-boson fiducial cross sections in the diphoton decay channel using 139
  fb$^{-1}$ of $pp$ collision data at $\sqrt{s}$ = 13 TeV with the ATLAS
  detector}}, .

\bibitem{Sirunyan:2018ouh}
{\bf CMS} Collaboration, A.~Sirunyan et~al., {\it {Measurements of Higgs boson
  properties in the diphoton decay channel in proton-proton collisions at
  $\sqrt{s} =$ 13 TeV}},  {\em JHEP} {\bf 11} (2018) 185,
  [\href{http://arxiv.org/abs/1804.02716}{{\tt arXiv:1804.02716}}].

\bibitem{Aad:2019vnb}
{\bf ATLAS} Collaboration, G.~Aad et~al., {\it {Search for electroweak
  production of charginos and sleptons decaying into final states with two
  leptons and missing transverse momentum in $\sqrt{s}=13$ TeV $pp$ collisions
  using the ATLAS detector}},  {\em Eur. Phys. J. C} {\bf 80} (2020), no.~2
  123, [\href{http://arxiv.org/abs/1908.08215}{{\tt arXiv:1908.08215}}].

\bibitem{Aad:2019byo}
{\bf ATLAS} Collaboration, G.~Aad et~al., {\it {Search for direct stau
  production in events with two hadronic $\tau$-leptons in $\sqrt{s} = 13$ TeV
  $pp$ collisions with the ATLAS detector}},  {\em Phys. Rev. D} {\bf 101}
  (2020), no.~3 032009, [\href{http://arxiv.org/abs/1911.06660}{{\tt
  arXiv:1911.06660}}].

\bibitem{ATLAS:2020ckz}
{\bf ATLAS} Collaboration, {\it {Search for chargino-neutralino pair production
  in final states with three leptons and missing transverse momentum in
  $\sqrt{s}$ = 13 TeV p-p collisions with the ATLAS detector}}, .

\bibitem{Alwall:2014hca}
J.~Alwall, R.~Frederix, S.~Frixione, V.~Hirschi, F.~Maltoni, O.~Mattelaer,
  H.~S. Shao, T.~Stelzer, P.~Torrielli, and M.~Zaro, {\it {The automated
  computation of tree-level and next-to-leading order differential cross
  sections, and their matching to parton shower simulations}},  {\em JHEP} {\bf
  07} (2014) 079, [\href{http://arxiv.org/abs/1405.0301}{{\tt
  arXiv:1405.0301}}].

\bibitem{Boehm:2013jpa}
C.~Boehm, M.~J. Dolan, and C.~McCabe, {\it {A Lower Bound on the Mass of Cold
  Thermal Dark Matter from Planck}},  {\em JCAP} {\bf 08} (2013) 041,
  [\href{http://arxiv.org/abs/1303.6270}{{\tt arXiv:1303.6270}}].

\bibitem{Aad:2019vvf}
{\bf ATLAS} Collaboration, G.~Aad et~al., {\it {Search for direct production of
  electroweakinos in final states with one lepton, missing transverse momentum
  and a Higgs boson decaying into two $b$-jets in $pp$ collisions at
  $\sqrt{s}=13$ TeV with the ATLAS detector}},  {\em Eur. Phys. J. C} {\bf 80}
  (2020), no.~8 691, [\href{http://arxiv.org/abs/1909.09226}{{\tt
  arXiv:1909.09226}}].

\bibitem{Belanger:2014vza}
G.~B\'elanger, F.~Boudjema, A.~Pukhov, and A.~Semenov, {\it {micrOMEGAs4.1: two
  dark matter candidates}},  {\em Comput. Phys. Commun.} {\bf 192} (2015)
  322--329, [\href{http://arxiv.org/abs/1407.6129}{{\tt arXiv:1407.6129}}].

\bibitem{Nollett:2014lwa}
K.~M. Nollett and G.~Steigman, {\it {BBN And The CMB Constrain Neutrino Coupled
  Light WIMPs}},  {\em Phys. Rev. D} {\bf 91} (2015), no.~8 083505,
  [\href{http://arxiv.org/abs/1411.6005}{{\tt arXiv:1411.6005}}].

\bibitem{Heo:2015kra}
J.~H. Heo and C.~Kim, {\it {Light Dark Matter and Dark Radiation}},  {\em J.
  Korean Phys. Soc.} {\bf 68} (2016), no.~5 715--721,
  [\href{http://arxiv.org/abs/1504.00773}{{\tt arXiv:1504.00773}}].

\bibitem{Sabti:2019mhn}
N.~Sabti, J.~Alvey, M.~Escudero, M.~Fairbairn, and D.~Blas, {\it {Refined
  Bounds on MeV-scale Thermal Dark Sectors from BBN and the CMB}},  {\em JCAP}
  {\bf 01} (2020) 004, [\href{http://arxiv.org/abs/1910.01649}{{\tt
  arXiv:1910.01649}}].

\bibitem{Chu:2018qrm}
X.~Chu, J.~Pradler, and L.~Semmelrock, {\it {Light dark states with
  electromagnetic form factors}},  {\em Phys. Rev. D} {\bf 99} (2019), no.~1
  015040, [\href{http://arxiv.org/abs/1811.04095}{{\tt arXiv:1811.04095}}].

\bibitem{Aprile:2019xxb}
{\bf XENON} Collaboration, E.~Aprile et~al., {\it {Light Dark Matter Search
  with Ionization Signals in XENON1T}},  {\em Phys. Rev. Lett.} {\bf 123}
  (2019), no.~25 251801, [\href{http://arxiv.org/abs/1907.11485}{{\tt
  arXiv:1907.11485}}].

\bibitem{Agnes:2018ves}
{\bf DarkSide} Collaboration, P.~Agnes et~al., {\it {Low-Mass Dark Matter
  Search with the DarkSide-50 Experiment}},  {\em Phys. Rev. Lett.} {\bf 121}
  (2018), no.~8 081307, [\href{http://arxiv.org/abs/1802.06994}{{\tt
  arXiv:1802.06994}}].

\bibitem{Aprile:2019jmx}
{\bf XENON} Collaboration, E.~Aprile et~al., {\it {Search for Light Dark Matter
  Interactions Enhanced by the Migdal Effect or Bremsstrahlung in XENON1T}},
  {\em Phys. Rev. Lett.} {\bf 123} (2019), no.~24 241803,
  [\href{http://arxiv.org/abs/1907.12771}{{\tt arXiv:1907.12771}}].

\bibitem{GrillidiCortona:2020owp}
G.~Grilli~di Cortona, A.~Messina, and S.~Piacentini, {\it {Migdal effect and
  photon Bremsstrahlung: improving the sensitivity to light dark matter of
  liquid argon experiments}},  {\em JHEP} {\bf 11} (2020) 034,
  [\href{http://arxiv.org/abs/2006.02453}{{\tt arXiv:2006.02453}}].

\bibitem{Agnes:2018oej}
{\bf DarkSide} Collaboration, P.~Agnes et~al., {\it {Constraints on Sub-GeV
  Dark-Matter--Electron Scattering from the DarkSide-50 Experiment}},  {\em
  Phys. Rev. Lett.} {\bf 121} (2018), no.~11 111303,
  [\href{http://arxiv.org/abs/1802.06998}{{\tt arXiv:1802.06998}}].

\bibitem{Essig:2012yx}
R.~Essig, A.~Manalaysay, J.~Mardon, P.~Sorensen, and T.~Volansky, {\it {First
  Direct Detection Limits on sub-GeV Dark Matter from XENON10}},  {\em Phys.
  Rev. Lett.} {\bf 109} (2012) 021301,
  [\href{http://arxiv.org/abs/1206.2644}{{\tt arXiv:1206.2644}}].

\bibitem{Barak:2020fql}
{\bf SENSEI} Collaboration, L.~Barak et~al., {\it {SENSEI: Direct-Detection
  Results on sub-GeV Dark Matter from a New Skipper-CCD}},
  \href{http://arxiv.org/abs/2004.11378}{{\tt arXiv:2004.11378}}.

\bibitem{Raffelt:1996wa}
G.~Raffelt, {\em {Stars as laboratories for fundamental physics}: {The
  astrophysics of neutrinos, axions, and other weakly interacting particles}}.
\newblock 5, 1996.

\bibitem{Navarro:1996gj}
J.~F. Navarro, C.~S. Frenk, and S.~D. White, {\it {A Universal density profile
  from hierarchical clustering}},  {\em Astrophys. J.} {\bf 490} (1997)
  493--508, [\href{http://arxiv.org/abs/astro-ph/9611107}{{\tt
  astro-ph/9611107}}].

\bibitem{Vertongen:2011mu}
G.~Vertongen and C.~Weniger, {\it {Hunting Dark Matter Gamma-Ray Lines with the
  Fermi LAT}},  {\em JCAP} {\bf 05} (2011) 027,
  [\href{http://arxiv.org/abs/1101.2610}{{\tt arXiv:1101.2610}}].

\bibitem{Yuksel:2007ac}
H.~Yuksel, S.~Horiuchi, J.~F. Beacom, and S.~Ando, {\it {Neutrino Constraints
  on the Dark Matter Total Annihilation Cross Section}},  {\em Phys. Rev. D}
  {\bf 76} (2007) 123506, [\href{http://arxiv.org/abs/0707.0196}{{\tt
  arXiv:0707.0196}}].

\bibitem{PalomaresRuiz:2007eu}
S.~Palomares-Ruiz and S.~Pascoli, {\it {Testing MeV dark matter with neutrino
  detectors}},  {\em Phys. Rev. D} {\bf 77} (2008) 025025,
  [\href{http://arxiv.org/abs/0710.5420}{{\tt arXiv:0710.5420}}].

\bibitem{Klop:2018ltd}
N.~Klop and S.~Ando, {\it {Constraints on MeV dark matter using neutrino
  detectors and their implication for the 21-cm results}},  {\em Phys. Rev. D}
  {\bf 98} (2018), no.~10 103004, [\href{http://arxiv.org/abs/1809.00671}{{\tt
  arXiv:1809.00671}}].

\bibitem{Arguelles:2019ouk}
C.~A. Arg{\"u}elles, A.~Diaz, A.~Kheirandish, A.~Olivares-Del-Campo, I.~Safa,
  and A.~C. Vincent, {\it {Dark Matter Annihilation to Neutrinos: An Updated,
  Consistent \& Compelling Compendium of Constraints}},
  \href{http://arxiv.org/abs/1912.09486}{{\tt arXiv:1912.09486}}.

\bibitem{Campo:2018dfh}
A.~Olivares-Del~Campo, S.~Palomares-Ruiz, and S.~Pascoli, {\it {Implications of
  a Dark Matter-Neutrino Coupling at Hyper-Kamiokande}},  in {\em {53rd
  Rencontres de Moriond on Electroweak Interactions and Unified Theories}},
  pp.~441--444, 2018.
\newblock \href{http://arxiv.org/abs/1805.09830}{{\tt arXiv:1805.09830}}.

\bibitem{Bell:2020rkw}
N.~F. Bell, M.~J. Dolan, and S.~Robles, {\it {Searching for Sub-GeV Dark Matter
  in the Galactic Centre using Hyper-Kamiokande}},
  \href{http://arxiv.org/abs/2005.01950}{{\tt arXiv:2005.01950}}.

\end{thebibliography}\endgroup
}

\end{document}